\documentclass[12pt,letterpaper]{article}
\usepackage{multirow}
\usepackage{graphicx}
\usepackage{times}
\usepackage{xcolor}
\usepackage{longfigure}
\usepackage[superscript]{cite}

\usepackage[]{multibib}
\newcites{M}{References Methods}

\usepackage{setspace}
\newcommand{\zun}{\ensuremath{H_5O_2^+}}
\newcommand{\eig}{\ensuremath{H_9O_4^+}}
\newcommand{\ezun}{\ensuremath{H_{13}O_6^+}}

\title{The coupling of the hydrated proton to its first solvation shell}
\author{Markus Schr\"oder$^{1\ast}$, Fabien Gatti$^{2}$, David Lauvergnat$^3$,\\ Hans-Dieter Meyer$^{1}$, Oriol Vendrell$^{1\ast}$}

\begin{document}
\maketitle

\begin{center}
\noindent\normalsize{$^{1}$Theoretische Chemie, Physikalisch-Chemisches Institut, Universit\"at Heidelberg, Im Neuenheimer Feld 229, 69120 Heidelberg, Germany}\\
\normalsize{$^{2}$Universit\'e Paris-Saclay, CNRS, Institut des Sciences Mol\'eculaires d'Orsay,\\ 91405 Orsay, France}\\
\normalsize{$^{3}$Universit\'e Paris-Saclay, CNRS, Institut de Chimie Physique UMR8000,\\ 91405 Orsay, France}\\
~\\
\normalsize{$^\ast$Correspondence: markus.schroeder@pci.uni-heidelberg.de; oriol.vendrell@uni-heidelberg.de}
\end{center}

%\begin{abstract}
%\bf 
%    The radically different gas-phase infrared (IR) spectra of the Zundel (\zun)
%    and Eigen (\eig) cations indicate fundamentally different environments of
%    the solvated proton in its first solvation shell.
%    %
%    The question arises:
%    %
%    what is the least common denominator structure that explains the anharmonic
%    vibrational couplings and IR spectrum of the Zundel and Eigen cations, and
%    hence of the solvated proton?
%    %
%    Full dimensional quantum simulations of the IR spectrum of the Eigen and
%    Zundel cations demonstrate that two dynamical water molecules embedded in
%    the static environment of the parent Eigen cation constitute the fundamental
%    subunit featuring the spectral signatures and anharmonic couplings of the
%    solvated proton in its first solvation shell.
%\end{abstract} 

\section*{Abstract}
{\bf
The transfer of a hydrated proton between water molecules in aqueous
solution is accompanied by the large-scale structural reorganization of the
environment as the proton relocates, giving rise to the
Grotthus mechanism~\cite{mar99:601}.
The Zundel (\zun) and Eigen (\eig) cations are the main intermediate structures
in this process. They exhibit radically different gas phase infrared (IR)
spectra~\cite{ham05:244301,wol16:1131}, indicating fundamentally different
environments of the solvated proton in its first solvation shell.
The question arises:
is there a least common denominator structure that explains the IR spectra of
the Zundel and Eigen cations, and hence of the solvated proton?
Full dimensional quantum simulations of these protonated cations demonstrate
that two dynamical water molecules embedded in the static environment of the
parent Eigen cation constitute this fundamental subunit. It is sufficient
to explain the spectral
signatures and anharmonic couplings of the solvated proton in its first
solvation shell.
In particular, we identify the anharmonic vibrational modes that
explain the large broadening of the proton
transfer peak in the experimental IR spectrum of the Eigen cation, of which the
origin remained so far unclear.\cite{wol16:1131, yu19:1399}  
Our findings about the quantum mechanical structure of the first solvation shell
provide a starting point for further investigations of the larger protonated water
clusters with a second and additional solvation shells.}

\section*{Main}

%

%In solution, the proton motion is hence strongly coupled to other nuclear
%degrees of freedom.
Due to the complexity of the liquid phase, the infrared (IR) spectroscopy of
protonated water clusters in the gas phase opens a unique window to charaterize
and understand the elusive structural dynamics of this species.
For example, the IR spectrum of the Zundel cation (\zun) exhibits a
prominent Fermi resonance in the $\approx$ 1000~cm$^{-1}$ spectral region of the
shared proton mode due to its strong anharmonic coupling with a combination of
the wagging (water pyramidalization) and the oxygen-oxygen distance of the two
flanking water molecules~\cite{ven07:6918}.
This important feature, key to understanding the strong coupling of the shared
proton to its environment, could only be unambiguously measured following the
development of accurate messenger spectroscopy (based on Neon tagging) of the
gas-phase cation~\cite{ham05:244301}.  The theoretical assignment of this feature was a
computational \emph{tour de force} only possible due to the availability of a
high-quality potential energy surface~\cite{hua05:044308} in combination with
full-dimensional (15-dimensional) quantum dynamical calculations based on the
multi-configuration time-dependent Hartree (MCTDH)
approach~\cite{ven07:184302,ven07:184303,ven07:6918,ven08:4692,ven09:352}.

Recent measurements of the IR spectrum of the Eigen cation (\eig) reveal a
strong coupling between the proton transfer modes of the central hydronium unit
with the water molecules in its first solvation shell.
More importantly, they reveal strong shifts of the spectral position of the
proton transfer modes caused by the polarization through the tagging agent in
the second solvation shell~\cite{wol16:1131}.
The strong coupling with the first solvation shell leads to a large broadening
of the proton transfer band, now spanning about 500~cm$^{-1}$ and markedly
blue-shifted towards 2600~cm$^{-1}$ in comparison with the shared proton band of
the Zundel cation.
The unambiguous characterization of this very broad band has remained a
long-standing challenge and open question~\cite{wol16:1131, yu19:1399}.

In this paper, we simulate and assign the linear absorption spectrum of the Eigen cation
%both pristine and perdeuterated, 
in the spectral range 0-4000~cm$^{-1}$ using full-dimensional (33D) quantum
dynamical calculations. Our spectra are in excellent agreement with the
available messenger-tagging spectra in the full spectral range~\cite{wol16:1131}.
We compare the full spectrum of the Eigen cation with those calculated with
frozen subsets of degrees of freedom all the way down to a dynamical Zundel
subunit embedded in the static scaffold of the remaining Eigen cation.  This analysis reveals that the
underlying coupling mechanism of the solvated proton with its first solvation
shell is strikingly similar in both the Zundel and Eigen forms: a dynamical
subunit formed by two water molecules and a proton is the least common
denominator structure that reproduces the spectrum and anharmonic mode couplings
of the Zundel and Eigen forms depending on the conformation of its static
environment.
Along this analysis, we confirm existing assignments of various peaks in
{\eig},~\cite{wol16:1131, yu17:121102, yu17:10984, ess18:798} and assign
hitherto unknown features in the low frequency region, where no experimental
data is currently available.

\subsection*{IR spectrum of the Eigen cation}

\begin{figure}[!t]
  \centering
  \includegraphics[width=0.7\textwidth]{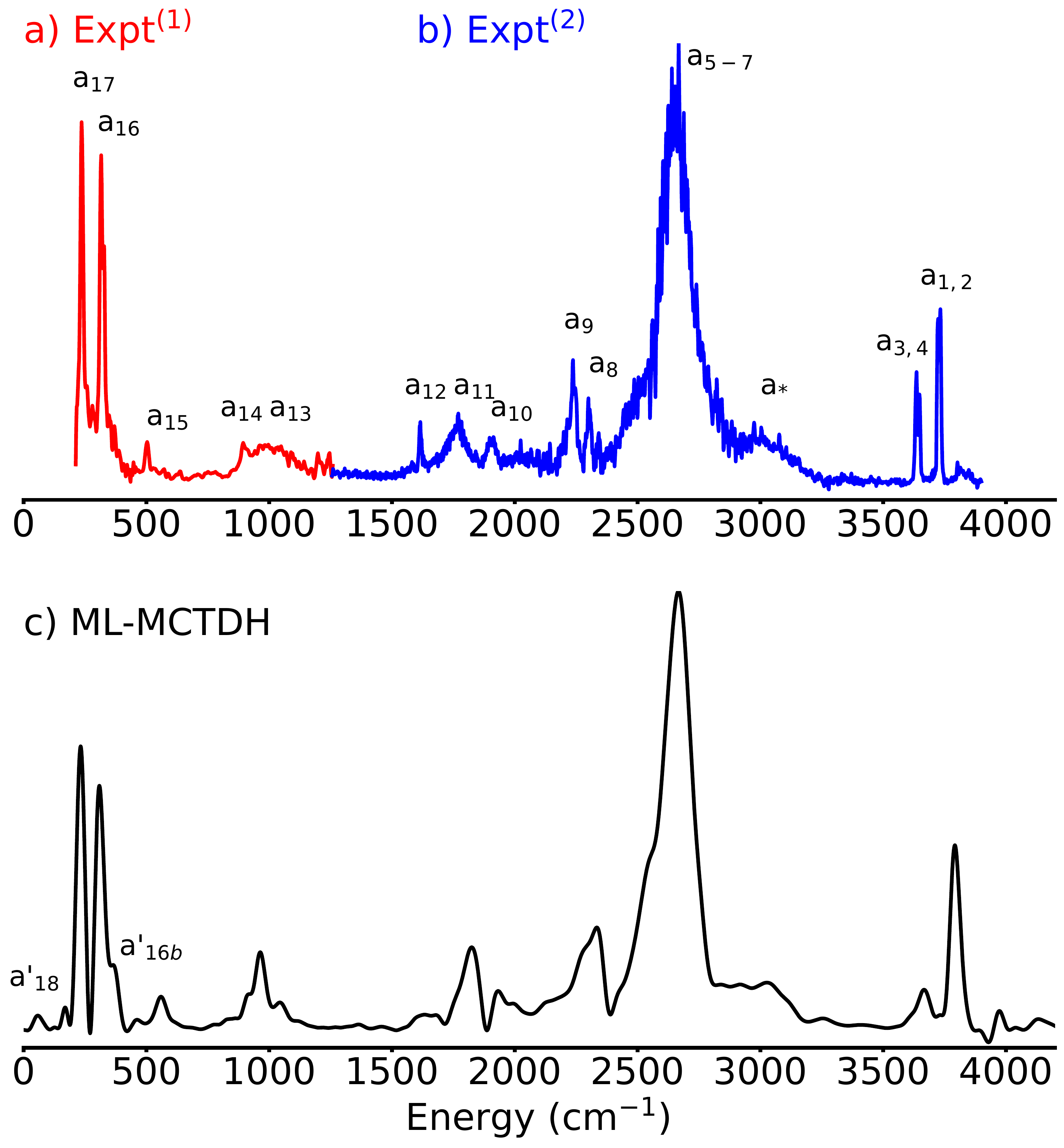}
  \caption{Absorption spectrum of the Eigen Cation H$_9$O$_4^+$. 
  a) Experimental spectrum from Ref.\ \citenum{ess18:798}, 
  b) Experimental spectrum from Ref.\ \citenum{wol16:1131}, 
  c) Calculated spectrum (red-shifted 70 cm$^{-1}$ to match experimental line positions).
  The assignments of the peaks follow the nomenclature of Refs.~\citenum{wol16:1131,ess18:798}
  and are discussed in Table~1 of the supporting material.
  \label{fig:hspectrum}}
\end{figure}
Figure\ \ref{fig:hspectrum} shows the calculated absorption spectrum of the
perprotio Eigen cation
%using a full-dimensional wavefunction and Hamiltonian of the Eigen cation
in comparison with the experimental spectra from Refs.\ \citenum{ess18:798} and\ \citenum{wol16:1131}.
The calculated IR spectra are based on a 33D quantum
mechanical description of the Eigen cation. Such simulations could be achieved only after the
unique combination of recent developments in our groups; They constitute the
largest quantum wavepacket simulations of a flexible molecular system
using a general potential energy surface and curvilinear coordinates reported to date.
Details of the 33-dimensional quantum-dynamical calculations including the
construction of the kinetic~\cite{ndo13:204107} and potential~\cite{sch20:024108} energy operators, and
the wavefunction propagations with the multilayer MCTDH
 method~\cite{wan03:1289,man08:164116,ven11:044135}, are provided as supporting information.

The calculated spectrum is red-shifted by 70 cm$^{-1}$ to match the main
features of the experimental spectrum. It is obtained as the average over the
spectra corresponding to the three polarization directions of light with respect
to the molecular frame, thus considering the random orientation of the molecules
in the experiment (see Methods and extended data for details).

The overall agreement of calculated and experimental spectra is very good
although the resolution of the calculated spectrum is approximately 30 cm$^{-1}$
and limited by the 1~picosecond duration of the dipole-dipole correlation
function. The calculated peak positions are listed in Table 1 of the supporting
material alongside with experimental results and assignments. In particular,
the substructure of the broad proton transfer band and practically all features
of the spectrum are reproduced in comparison with the tagging-agent IR measurement.
Our simulations thus further support the interpretation that
(i) the spectra in Refs.~\citenum{wol16:1131} and \citenum{ess18:798} correspond
to the triply-coordinated hydronium form of {\eig} stoichiometry,
and (ii) that the $D_2$ tagging-agent negligibly alters the spectrum
of {\eig$\cdot D_2$}~\cite{wol16:1131,ess18:798} compared to {\eig}.

%
%Interestingly, the O-H bending 
%of the ligand water at 1630 cm$^{-1}$ (marked $\ddag$) is significantly populated while 
%this signal is not visible in the absorption spectrum.

\subsection*{Deconstructing the broad proton transfer band}

\begin{figure}[!ht]
  \centering
  \includegraphics[width=0.8\textwidth]{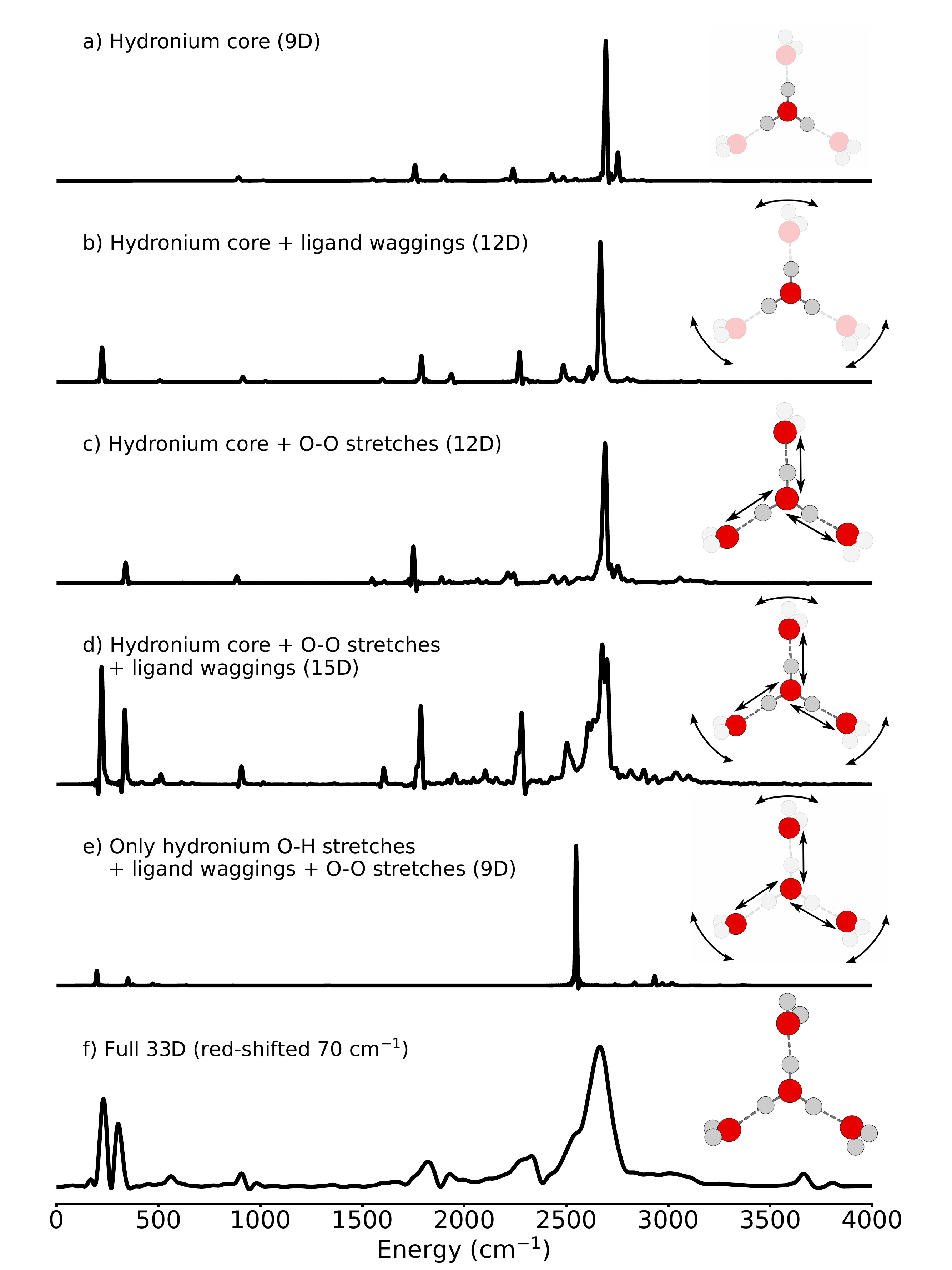}
  \caption{Spectra obtained with the z-component of the dipole moment surfaces for
  various reduced models by freezing modes to equilibrium positions for H$_9$O$_4^+$. 
  Correlation time in panels a-e): 2000 fs, panel f): 1000 fs.
  %\textcolor{red}{oriol: in c) and d) the arrows along the hydrogen bond could be crossed
  %by arrows indicating the perpendicular motions. This will make it visually different from e)
  %and more clear for the reader.}\textcolor{blue}{ markus: the arrows are ment to indicate the O-O stretch, not the hydrogen. When the perpendicular motion is switched off the hydrogens are grayed.}
  \label{fig:envH}}
\end{figure}

The key to understanding the anharmonic couplings of the proton transfer modes
to their first solvation shell lies in characterizing the broadening and
composition of the main proton transfer band in pristine H$_9$O$_4^+$: This
feature carries most of the IR intensity related to the coupled motions of the
central proton stretching modes.
We deconstruct the formation of this band by first freezing all modes of the
Eigen cation, except those of the hydronium core, to their equilibrium
positions, and then by successively bringing back the environment.
The spectra obtained in this way are shown in Fig.~\ref{fig:envH}. They
correspond to the z-component of the dipole moment (the polarization is aligned
with one of the hydronium hydrogen bonds), since this is the component
responsible for the largest response of the proton transfer modes. 
Freezing specific coordinates is achieved by removing all differential operators
of a frozen coordinate from the Hamiltonian in a Hermitian way and by fixing their position to the
corresponding expectation value in the ground vibrational state of the
full-dimensional system.

%We restrict the discussion to the z-component of the dipole moment here as it contains 
%all features of the broadening as demonstrated above. The $z$-component 
%points in the direction of one of the three "arms" of the Eigen cation such that the asymmetric O-H-stretch (-++) can be excited.

%The spectra obtained with the dipole z-component for
%various reduced models are depicted in Fig. \ref{fig:envH}, starting with the the spectrum of the
%hydronium core (9D) with all other modes frozen to equilibrium positions in
%panel a).
The IR spectrum of the hydronium core embedded in the frozen environment (cf.
Fig. \ref{fig:envH} a)) has a very simple structure. The vibrational eigenstates
corresponding to the two sharp peaks near 2700 cm$^{-1}$ were obtained by full
diagonalization explicitly: The dominant peak corresponds to the proton transfer
mode, whereas the smaller structure corresponds to an out-of-plane excitation of
the central hydrogen atoms.
The peaks near 1800 and 2300 cm$^{-1}$ correspond to other modes of the
hydronium core also seen in the full spectrum and agree with the assignments in
Refs.\ \citenum{wol16:1131, yu17:121102, yu17:10984, ess18:798}.

%Since all other modes are frozen, all peaks in Fig. \ref{fig:envH} a) must belong to the states 
%of the hydronium core. This justifies the assignments of peaks near 1800 and 2300 cm$^{-1}$ as states of
%the hydronium core in Refs.\ \cite{wol16:1131, yu17:121102, yu17:10984, ess18:798}.
%One observes two sharp peaks near 2700 cm$^{-1}$. The respective states were
%calculated with the improved relaxation algorithm of MCTDH\cite{mey06:179,dor08:224109}.
%The dominant peak is the proton transfer mode, the other is mainly an 
%out-of-plane excitation of the central hydrogen atoms where zero-order
%quantum numbers cannot be assigned due to the complex nature of the wavefunction. 
%Energetically it could be a triple excitation of an out-of-plane motion, i.e., 
%a combination of hydronium wagging and umbrella motions. 

\begin{figure}[!t]
  \centering
  \includegraphics[width=0.7\textwidth]{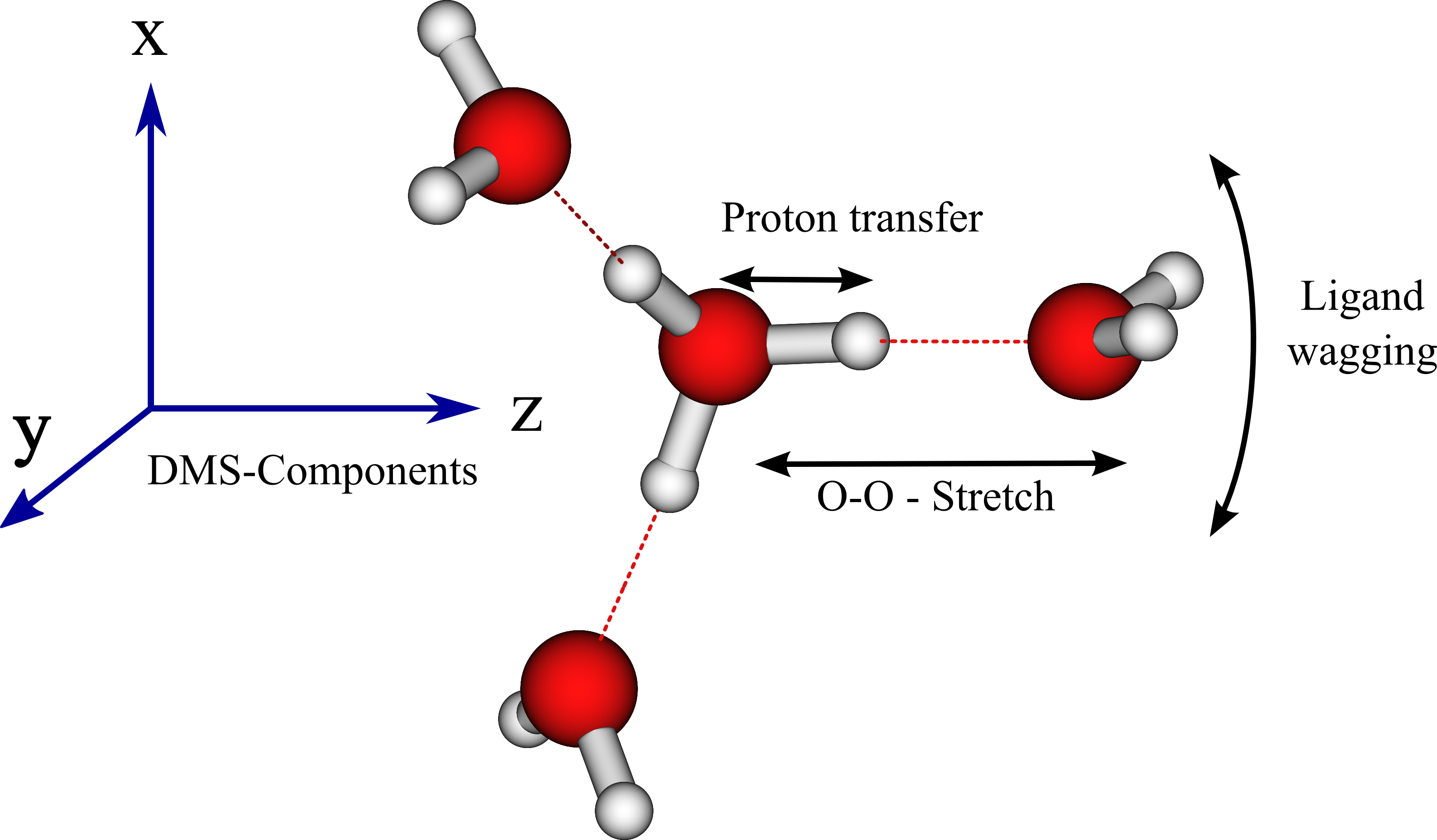}
  \caption{Illustration of ligand wagging, proton transfer and O-O stretching motion of H$_9$O$_4^+$ exemplary 
  in one of the three ``arms'' of the cation. Note that in the ligand wagging motion only the two hydrogen of the outer water molecules move as indicated by the arrow.
  \label{fig:wag}}
\end{figure}

Adding either wagging modes of the outer water molecules, Fig. \ref{fig:envH}
b), or O-O distances, Fig. \ref{fig:envH} c), leads to the appearance of their
fundamental modes in the spectrum (Illustrated in Fig. \ref{fig:wag}).  In the
latter case, some peaks on the low energy shoulder of the main proton transfer
peak gain some intensity. Moreover, with the inclusion of the O-O stretching
coordinates, two small peaks appear at 2300~cm$^{-1}$ correlating with a$_8$ and
$a_9$ in the full dimensional spectrum.  Apart from this, the overall structure
of the spectrum changes only slightly. In particular, there is no significant
broadening of the proton transfer peak.

More complex spectral features emerge when adding both solvation-shell water
wagging modes and O-O distances together (Fig.~\ref{fig:envH} d). Now, the spectrum
is not the simple sum of the previous two panels and cannot be explained by the
fundamental modes of the involved coordinates alone.
The broad proton transfer band centered at 2700 cm$^{-1}$ is now composed of at
least four separate contributions with significant intensity.
%for the z-component of the  dipole moment surface alone.  Only one of them can
%be attributed to the hydronium O-H stretch due to symmetry.
Two of those peaks, contributing to the low energy shoulder of the central peak
at approximately 2600 cm$^{-1}$, have gained significant intensity.
Finally, a peak slightly above 2500 cm$^{-1}$ gains significant intensity as
well. This structure coincides with the spectral position of the low energy
shoulder of the broad band in the full spectrum.
%\textcolor{red}{oriol: would be good if we could formulate this paragraph
%a bit more concisely by specifying the nature of the states we are discussing.
%Do we have this information?}
%\textcolor{blue}{markus: Not really. This is what we look into in
%the Zundel@Eigen section. But for most peaks I cannot really state quantum numbers}

%broadening at the low energy
%shoulder such that the spectrum almost resembles the shape of the spectrum in the full-dimensional system, shown in panel f). 
%Note that the spectrum of the full-dimensional system in panel f) has less spectral resolution due to shorter propagation time.

In the spectra in panels a) to d), the hydronium core retains its full mobility.
The question arises, whether only proton displacements parallel to the hydrogen
bonds are important, or whether displacements perpendicular to the hydrogen
bonds also contribute to the main proton-transfer band.
These perpendicular displacements span the hydronium bending, wagging, and
pyramidalization modes.
Freezing the perpendicular displacements of the hydronium protons (panel e)) has
dramatic consequences. The spectrum is now dominated solely by the proton
transfer peak. Peaks of the ligand wagging and O-O stretching fundamentals are
again visible with low intensity at low energies, as well as peaks at
approximately 3000 cm$^{-1}$ that are combinations of proton transfer, ligand
wagging, and O-O stretch modes. However, the inability of the three central
protons to move perpendicular to the proton transfer directions has largely
suppressed their coupling with the first shell of ligand water molecules.  
Crucially, no broadening of the proton transfer peak is present, as opposed to
the spectrum in panel d).
This leads to the conclusion that the
% unresolved (both in experiment and in theory)
vibrational eigenstates spanning the broad proton transfer band correspond to
combinations and overtones of the proton transfer modes with O-O stretch
displacements, hydronium bending and hydronium wagging, and ligand waggings,
whereby none of those coupled hydronium and environment modes can be removed.
%\textcolor{red}{oriol: no ligand bendings involved at this point, right?}
%
A full characterization of the vibrational eigenstates in terms of quantum
numbers of some basis of uncoupled vibrational modes is currently out of reach
due to the very high density of vibrational states in the spectral region of the
band and the high dimensionality of the problem.

\subsection*{The dynamical Zundel subsystem}

%\begin{figure}[!ht]
%  \centering
%  \includegraphics[width=0.4\textwidth]{fig/eigen_zundel}
%  \caption{Illustration of the Zundel subunit. Frozen parts are grayed out.
%  Also the two outer OH-stretching modes pointing to the left water (former hydronium core) are frozen.
%  Furthermore, frozen additionally for the 9D model are: internal coordinates of the right water molecule %as well as its rotation around the O-O axis
%  and rocking (wagging included).  
%  \label{fig:eigenzundel}}
%\end{figure}
%

\begin{figure}[!t]
  \centering
  \includegraphics[width=0.8\textwidth]{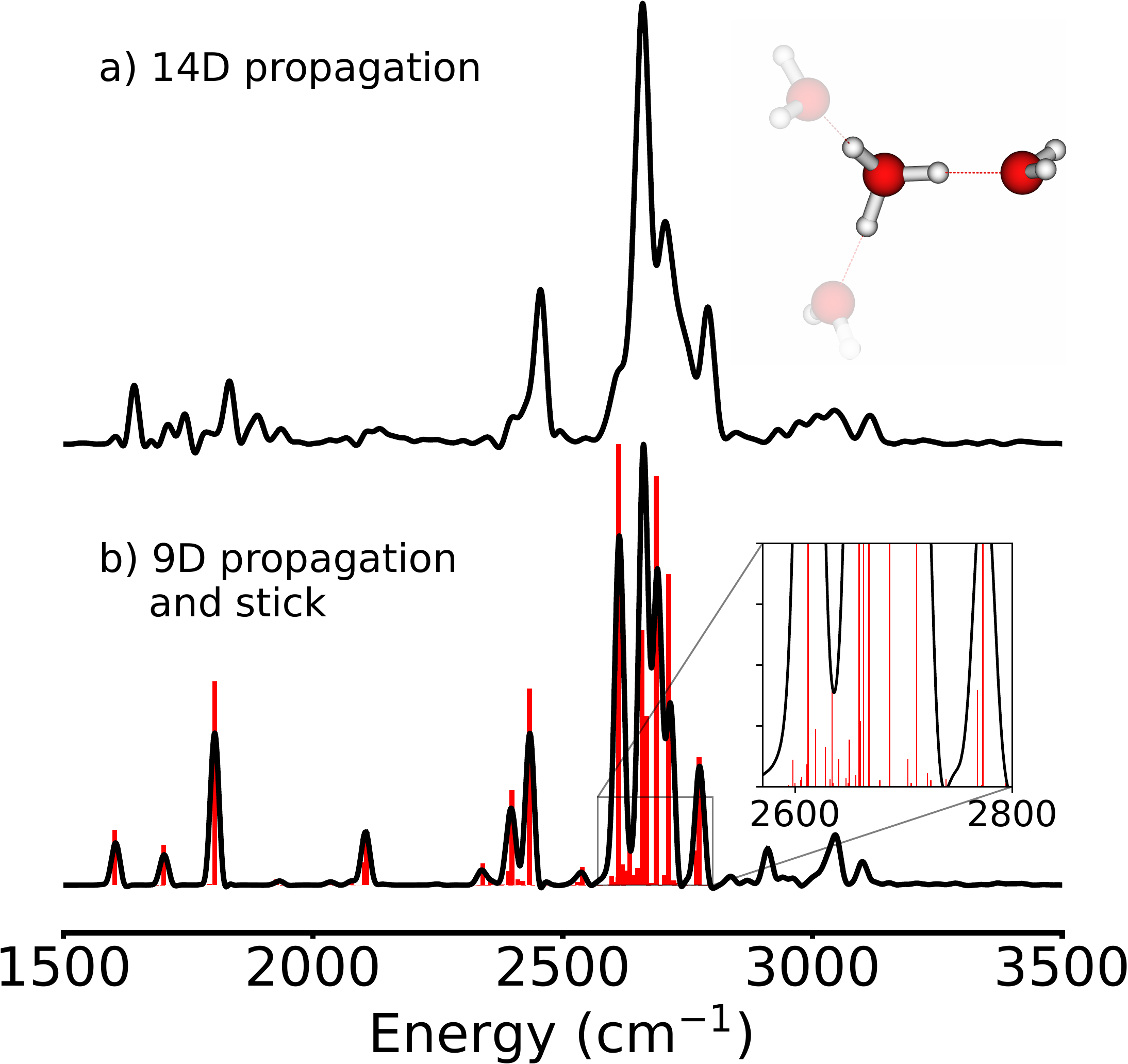}
  \caption{Spectra obtained with the z-component of the dipole moment surfaces for
  a reduced Zundel-like model obtained by freezing modes to equilibrium positions for H$_9$O$_4^+$  
  a) obtained with a dipole-dipole-correlation function of 2000 fs using a 14D model,
  b) obtained with a dipole-dipole-correlation function of 2000 fs using a 9D model,
 and obtained as a stick spectrum using eigenstates. 
  \label{fig:tinyzundel_H}}
\end{figure}

We have deconstructed the main proton transfer band. It originates from
the anharmonic couplings of the proton transfer modes with perpendicular modes
of the central hydronium and modes involving the O-O stretchings and waggings of
the three surrounding water ligands.
The question now arises: Are the three water molecules in the first solvation
shell of the Eigen cation necessarily involved in explaining the coupling
mechanism, spectral position, and width of the main proton-transfer band?
Alternatively, can
a smaller dynamical subunit completely account for the properties of the first
solvation shell of the solvated proton?
The hydronium cation (H$_3$O$^+$) can be discarded as the least common denominator subunit
by comparing Figs.~\ref{fig:envH} a) and f).
%
%To further reduce the complexity and investigate the nature of the signals in
%Fig.  \ref{fig:envH} d) that lead to the broadening of the proton transfer peak
%we removed a number of degrees of freedom from the model used in Fig. 
%\ref{fig:envH} d). Effectively the model used in the following coincides with a
%Zundel subunit forming one "arm" of the Eigen cation, embedded in a frozen
%Eigen environment. Here we explicitly modeled seven DOF of the hydronium core,
%one of which being a proton transfer coordinate of the  Zundel subunit of the
%Eigen cation, the others describing the six remaining hydronium bending,
%wagging/umbrella and frustrated rotation motions.  Furthermore one O-O-distance
%and one ligand wagging mode, both within the respective arm, were included in
%the model, cf. Fig.\ \ref{fig:eigenzundel}  
%

Instead, we consider one Zundel cation subunit dynamically (Zundel@Eigen, 14
coordinates) and freeze all internal, angular and relative coordinates of the
two other water ligands to their equilibrium positions (cf.
Fig.~\ref{fig:tinyzundel_H}), as well as the two free-standing hydronium O-H
stretches, as those do not interact with their immediate environment dynamically
any more.
For comparison, we also consider a reduced version of the Zundel@Eigen cation
where the rocking, relative water rotation and internal modes of the external
water are also frozen, thus yielding a 9-dimensional system for which the lowest
250 eigenstates can be computed with the improved relaxation
algorithm (ticks in Fig.~\ref{fig:tinyzundel_H}b))~\cite{mey06:179,dor08:224109}.

%With this reduction we obtained a 9D reduced model that is small enough to be able to compute the 250 lowest 
%eigenstates using the improved relaxation algorithm of the Heidelberg MCTDH package. 
%The resulting stick spectrum is depicted in Fig. \ref{fig:tinyzundel_H} in comparison with the dynamical
%spectrum obtained with the same reduced model.

The IR spectrum of the dynamical Zundel@Eigen cation
% embedded in the rest of the static Eigen cation scaffold
is strickingly similar to the full Eigen cation spectrum, as seen in
Fig.~\ref{fig:tinyzundel_H}. The main proton transfer band presents a comparable
broadening and is centered at the same frequency. Other flanking peaks appear at
the correct positions as well.
The analysis of 1D and 2D probability densities of
the calculated eigenstates of the 9-dimensional model reveal that the
vibrational states that participate in this band are complex combinations and
overtones of the same
% with similar excitation quanta to those
vibrational coordinates previously found to contribute to the broadening of the proton transfer
band in the Eigen cation.
%
%\textcolor{red}{oriol: do we have this analysis in some form, e.g. as supporting
%material, that can substantiate our claim?}
%
Just pulling the external water molecules by about {0.5~\AA} away from the central
hydronium, while leaving them frozen, results in a shift of the proton transfer
band to the red by about 600~cm$^{-1}$ (cf. supporting material), indicating the
extreme sensitivity of the position of this band to the polariztion by the first
solvation shell of water molecules.  Pulling them further to infinity leaves the
bare Zundel cation with its proton transfer band red-shifted by about
1600cm$^{-1}$ compared to the Eigen cation~\cite{ven07:6918}.

Based on these observations, we argue that \emph{two} protonated water
molecules, nominally the Zundel subunit, constitute the \emph{dynamical} least
common denominator structure explaining the anharmonic couplings and spectral
signatures of the solvated proton in its first solvation shell.
This statement does not concern the relative population of the Zundel and Eigen
structures in solution, which has been investigated separately by Marx and
collaborators using path integral thechniques, cf. Ref.~\citenum{mar99:601}.
%, which is a well-studied question and has been the subject of continued
%experimental and theoretical efforts [CITATIONS].
%
%Instead, the Zundel subunit features, depending on its environment, the
%fundamental spectroscopic signatures and anharmonic couplings of the excess
%proton in its isolated Zundel or Eigen forms.
%

%
In isolation, the main shared-proton peaks in the Zundel cation are strongly
red-shifted compared to the Eigen cation. The shared-proton motion strongly
couples to the wagging (pyramidalization) of the two water molecules and to the
O-O stretching mode, and results in the well-characterized Fermi resonance
doublet centered at about 1000 cm$^{-1}$.\cite{ven07:184302,ven07:184303,ven09:234305,ven09:034308}
Embedded in the potential of two flanking, frozen water molecules,
the Zundel@Eigen subsystem features its proton transfer band at the same spectral
position as the full-dimensional Eigen cation, i.e. blue-shifted to
about 2600 cm$^{-1}$ because the shared proton is now much closer to the central water molecule.
The broadening of the proton transfer band in the Zundel@Eigen and Eigen cations
is strikingly similar. Our simulations demonstrate that the same set of vibrational
coordinates and corresponding combined excitations are responsible for the
strong coupling of the shared proton to the rest of the scaffold in the
Zundel~\cite{ven07:184302,ven07:184303,ven09:234305,ven09:034308}, Zundel@Eigen
and Eigen cations. These effects are strongly cooperative as opposed to
additive.  These relevant coordinates are the hydronium O-H bending and wagging modes,
the ligand water wagging modes, and the O-O hydrogen bond stretching mode.

\subsection*{Conclusions and outlook}

This work has provided a first set of full-dimensional quantum simulations of
the Eigen cation 
based on flexible, curvilinear coordinates and a very accurate potential energy
representation.
The simulated IR spectra cover the chemically relevant spectral range between 0
and 4000~cm$^{-1}$ with one single time-propagation of a highly correlated
multiconfigurational wave function.  The spectra extend below the smallest
frequency accessible experimentally using ion tagging techniques, and reveal the
signatures of very low frequency, global vibrational modes.
Both the Zundel and Eigen cations feature very prominent spectral features
related to the anharmonic couplings of the hydrated proton with its first solvation
shell. In the Zundel cation, a strongly red-shifted double
peak\cite{ven07:184303,ven09:034308} originates from the fundamental vibration
of the equally shared proton at about 1000~cm$^{-1}$.  This doublet is a Fermi
resonance that involves the wagging modes of the flanking water molecules as
well as the hydrogen-bond O-O stretching.  The isolated Eigen cation, instead,
features a very broad band at 2600~cm$^{-1}$ with little resemblance to the
shape and position of the Zundel's double peak.
Nonetheless, a careful analysis reveals that similar anharmonic couplings compared to the Zundel form
are
involved in the broad Eigen cation band, namely the hydronium and ligand
waggings to the largest extent, combined with hydrogen-bond stretchings.
Indeed, the hydronium waggings are crucial to the coupling mechanism: freezing
the central hydronium waggings in a flexible first solvation shell results in a
simpler IR spectrum than when considering a fully flexible hydronium in a frozen
environment (cf. Figs.~\ref{fig:envH}a and e).

Based on these results and observations, we arrive at a key insight: two
dynamical water molecules and a proton, i.e. a Zundel subunit embedded in the
remaining frozen scaffold of the Eigen cation, presents all anharmonic couplings
and spectral signatures of the fully flexible Eigen cation in the region of the
main proton-transfer band.
For this effect, it is sufficient that two frozen, hydrogen-bond acceptor water
molecules polarize the dynamical Zundel subunit that constitutes the proton's first
solvation shell.
This finding, backed by our full quantum-mechanical approach, is suggestive of
picturing the Eigen cation as three overlapping and strongly polarized Zundel
subunits in the spirit of the classical special pair models of the solvated
proton~\cite{mar_08_9456, kul_13_29}. 
In turn, one can similarly anticipate a similar relation in the second solvation
shell, where the extended Zundel cation {\ezun}
can be pictured as two superimposed and dynamically overlapping Eigen
cations.
%
%The unique computational and theoretical developments that have enabled the
%simulation of the Eigen cation will be key in approaching even larger and more
%complex systems.
%
Establishing these nested structural and dynamical relations on the basis of
full-dimensional quantum dynamics is an important direction for future work.
This is now enabled by the unique computational and theoretical developments
reported in this work, which will be decisive when approaching even larger and
more complex systems.

%\bibliographystyle{naturemag}
%\bibliography{refs}

\section*{Methods}

\subsection*{High-dimensional quantum dynamics}

The full-dimensional (33 vibrational degrees of freedom) quantum dynamical
description of the IR spectrum of the Eigen cation requires the combination of
various technologies that have been developed and integrated into the software
packages maintained in our research groups. 
These technologies relate to the three main obstacles that stand on the way
towards a full quantum dynamical description of anharmonically coupled,
flexible, and high-dimensional vibrational problems.

(i) % TANA
Describing flexible and anharmonic systems, e.g. with several equivalent minima
in their potential energy surface (PES), requires the use of chemically
meaningful coordinates such as bond lengths, bond angles, and dihedral angles.
The use of adequate coordinates enormously facilitates the numerical
representation and convergence of the vibrational wavefunctions in
high-dimensions.
The price to pay, though, is the very lengthy and complicated expression for the
corresponding kinetic energy operator (KEO). For the Eigen cation, the exact,
analytic KEO has a total of 4370 terms and its manual derivation becomes
\textit{de facto} intractable.
Some of us and others have therefore developed a completely systematic method to set up the KEO for a specific family of internal molecular coordinates: the polyspherical
coordinates~\citeM{gat98:8804,gat99:7225,gat09:1}.
This method is implemented in the TANA software, which provides analytic
expressions of the kinetic energy operator in a machine readable
format~\cite{ndo13:204107}\citeM{ndo12:034107,tana-tnum:package}.
Very importantly, TANA also provides numerical library routines to perform
forward and backward transformations between the Cartesian coordinates of the
atoms and the internal coordinates of the molecule, which are needed when
setting up the potential energy operator in these internal coordinates.

(ii) % Wavefunction
The second obstacle is the so-called ``curse of dimensionality'' for
representing and storing the wavefunction of the system: the number of possible
quantum states of the system (e.g. given as the amplitudes on quadrature points in
coordinate representation) grows exponentially with the number of physical
coordinates. Without an efficient data reduction scheme one would be limited to
model up to about six internal degrees of freedom of a molecule, corresponding
to about four atoms (neglecting rotations). To overcome the curse of
dimensionality, the state vector needs to be stored and processed in a very
compact form.
To this end, we employ the multi-layer multi-configuration time-dependent
Hartree algorithm,~\cite{wan03:1289, man08:164116, ven11:044135}\citeM{mey12:351,wan15:7951}   
which represents the wavefunction as a hierarchical Tucker
tensor-tree~\citeM{tuc66:279,hac09:706,gra11:291}.

(iii) % Potential
The solution of the time-dependent Schr\"odinger equation within this tensor
format requires that also the system Hamiltonian is expressed in a matching
form. This can be, e.g., a sum of products of low-dimensional operators.  The
KEO in polyspherical coordinates always consists of sums of products of
elementary functions and derivatives of single coordinates \citeM{gat09:1} 
(this is one of the main advantages of the polyspherical coordinates) and needs
not be discussed further here.
A more challenging task is to express the PES and, if needed, other surface-like
operators such as dipole moment surfaces (DMS), in a matching format.  The PES
and DMS are usually made available as separate  software libraries, and are
often defined in the Cartesian coordinates of the atoms~
\cite{yu17:121102,yu17:10984}\citeM{duo17:3782}.
Most applications in our groups have relied until recently on the transformation
of the PES into a Tucker format with the so-called Potfit algorithm
\citeM{bec00:1,jae96:7974,jae98:3772}, and its hierarchical multi-layer 
variant\citeM{ott14:014106}.
This algorithm suffers from the curse of dimensionality because ultimately it
requires a full representation of the primitive product grid in configuration
space. Modifications of Potfit have been developed over the years to partially
overcome this difficulty, \citeM{pel13:014108,sch17:064105,ott18:116} making it
possible to work with about 9 to 15 coordinates. This is clearly insufficient to
approach a system of the size of the Eigen cation.  
A more recent development in surface re-fitting uses the so-called canonical
tensor decomposition\citeM{hit27:164} (CP), also called PARAFAC or CANDECOMP in
the literature\citeM{har70:1, car70:283}. Within the canonical format,
orthogonality restrictions on the basis functions are relaxed such that a much
more compact tensor representation can be achieved, at the cost however, that
the fit is much harder to obtain. This is usually achieved using an alternating
least squares (ALS) algorithm that iteratively improves an initial guess tensor.
The ALS algorithm in the original form requires to perform high-dimensional
integrals as well. In a recent publication\cite{sch20:024108} Monte-Carlo
integrations are used to perform the integrals. This not only mitigates the
curse of dimensionality but also allows for importance sampling such that low
energy regions of the potential (where the wavefunction resides) can be fitted
with elevated accuracy. This development has opened the path to obtain global
but compact surface fits in a tensor format of high-dimensional potentials.

In essence, we developed and combined 
three technologies for the first time to be able tackle such a high-dimensional 
problem as the 33-dimensional Eigen cation: 1) the TANA software to obtain the KEO
and to provide the coordinate transformations for the PES fitting; 2) PES fitting 
into a canonical tensor format using a Monte-Carlo version of the ALS algorithm; 
and 3) the multi-layer MCTDH algorithm to solve the time-dependent Schr\"odinger equation. 
In the present contribution we have used the highly accurate, full-dimensional PES and DMS
provided by Yu and Bowman\cite{yu17:121102,yu17:10984}\citeM{duo17:3782}.
The surfaces were re-fitted into a canonical tensor format using 2048 terms for
the PES and 1024 terms for each of the three components of the DMS,
respectively. 
%The KEO created with the TANA software consists of 4370 terms.

\subsection*{Calculation of IR Spectra}

The linear absorption spectra that are compared to the experimental spectra are computed as averages of 
the spectra resulting from the three dipole moment components for the $x$- $y$- and $z$-directions as
\begin{equation}
 I(\omega) = \frac13 (I_x(\omega) + I_y(\omega) + I_z(\omega))\, .
\end{equation}
The averaging  mimics the random orientational distribution of the molecule in the experiment. 
The single components also shown in some figures below 
are calculated as\cite{ven07:184303}
\begin{equation}
 I_j(\omega) \propto \omega \,{\rm Re} \int\limits_0^\infty dt \; \left< \Psi_{\mu_j} \right.\left| \Psi_{\mu_j}(t) \right> \exp(i (\omega + E_0/\hbar) t), ~~~~~~ j=x,y,z
 \label{eq:def_spect}
\end{equation}
where  $E_0$ is the ground state energy and  
\begin{equation}
 \left|\Psi_{\mu_j}  \right> = \mu_j  \left|\Psi_0  \right>~~~~~~ j=x,y,z
\end{equation}
is the vibrational ground state $\left|\Psi_0 \right>$ operated with $\mu_j$,  one component of the dipole operator. The time-dependent state
$\left| \Psi_{\mu_j}(t) \right>$ is obtained by solving the time-dependent Schr\"odinger equation with initial value  $\left|\Psi_{\mu_j} \right>$.

\subsection*{Assignments}

To assign modes to the peaks a number of test states that contain zero order excitations
in selected modes have been created and cross-correlated with the dipole operated and propagated 
ground state. 

The Fourier transform of the resulting cross correlation shows peaks only at frequencies 
where both, the test states and dipole operated ground states populate the same eigenstate.  
The cross-correlation-functions are defined as
\begin{eqnarray}
 C_{i,X}(t) = \left<\Psi_X\right.\left|\Psi_{\mu_i}(t)\right>  ~~~~~~ i=x,y,z
\end{eqnarray}
and $\left|\Psi_X\right> =X\left|\Psi_0\right>$ being the $X$-operated ground state with an operator $X$ as detailed below.
The Fourier transformed of the cross-correlation is given as
\begin{eqnarray}
 F_{i,X}(\omega) = \propto \,{\rm Re} \int\limits_0^\infty dt \;  \left<\Psi_X\right.\left|\Psi_{\mu_i}(t)\right> e^{i\left(\omega+E_0/\hbar\right) t} ~~~~~~ i=x,y,z,
 \label{eq:def_cross}
\end{eqnarray}
with $E_0$ being the ground state energy. Note that, other than for the absorption spectra, no frequency prefactor $w$ is multiplied to the spectrum.

The test-states $\left|\Psi_X\right> = X \left|\Psi_0\right> $ have been created by constructing the 
operator $X$ as linear combinations of
position operators of specific coordinates.  
This creates a linear combination of wavefunctions, with nodes in the respective modes,
hence resembling zero order excitations which mimic the action of the dipole moment 
surface but restrict the action only to the aforementioned modes. 
We use the notation $q(+++)$, $q(-++)$ and $q(0-+)$ for  $X$ in the test states. 
Here the $q$ indicate physical coordinates and the string of signs in brackets identifies
one of the three orthogonal linear combinations of the coordinates $q$ in the three 
'arms' A, B, and C of the Eigen cation (cf. Fig.\ S8, extended data), where specifically  
\begin{eqnarray}
 q(+++) & := & q_{\rm A} + q_{\rm B} + q_{\rm C} \nonumber \\
 q(-++) & := & -2q_{\rm A} + q_{\rm B} + q_{\rm C}  \label{eq:qtest}\\
 q(0-+) & := & -q_{\rm B} + q_{\rm C} \nonumber
\end{eqnarray}
(with the exception of label $q=$'$\theta$'$=(q_{\rm A}=\theta,q_{\rm B}=\varphi_{\rm AB}, q_{\rm C} = \varphi_{\rm BC})$,  
and q='b' describing the ligand O-H bending as a linear combination of two Jacobi coordinates b$_{\rm \{A,B,C\}}$ = -0.4 $r_{1{\rm,\{A,B,C\}}}$ + 0.3 $r_{2{\rm,\{A,B,C\}}}$. Similarly the symmetric O-H stretching of the ligands is described by q='$v^{(s)}$' with $v^{(s)}_{\rm \{A,B,C\}}$ = 0.3 $r_{1{\rm,\{A,B,C\}}}$ + 0.4 $r_{2{\rm,\{A,B,C\}}}$, while the asymmetric O-H stretching $v^{(a)}_{\rm \{A,B,C\}}$ = $\nu_{\rm \{A,B,C\}}$ is described by the Jacobi angle., cf. Table S1 of assignments and Table S2 and Fig. S9 of coordinate definitions in the extended data section).

Non-vanishing cross-correlations hence show the existence of non-vanishing overlap of 
the dipole operated state $\Psi_{\mu_i}$ and the test state  characterized by a 
linear combination of single mode excitations of character Eq.\ (\ref{eq:qtest}).

\subsection*{Kinetic energy operator}

As for the Zundel cation \cite{ven09:234305}, we adopted a mixture of Jacobi, Cartesian, and valence vectors. For each external molecule of water (in blue in Fig. S7, extended data), we use two Jacobi vectors: one from one hydrogen atom to the other and one from the middle of H$_2$ to the oxygen atom. The central oxygen atom is linked to the other oxygen atoms by three O-O valence vectors. The global $z$ Body-Fixed (BF) axis is parallel to $R_1^{BF}$, one of the O-O vectors. The groups $S^1$ and $S^2$ are gathered into two subsystems so that they have their own BF frame with the $z$ axis parallel to  $R_2^{BF}$ or  $R_3^{BF}$. The molecule at the top of Fig. S7 (extended data) is also gathered in one subsystem with the $z$ axis parallel to the H-H vector. The same is true for the other two molecules of water except that they define "subsubsystems" in $S^1$ and $S^2$. The three OH valence coordinates starting from the central oxygen atom are re-expressed in terms of Cartesian (and not spherical) coordinates to avoid  singularities in the kinetic energy operator (KEO).  

All the other vectors are parametrized by spherical coordinates in their BF frame. The rotation of each BF frame is parametrized by Euler angles. We follow the  conventions of the general formulation for polyspherical coordinates \citeM{gat04:3} that is implemented in the  TANA software\cite{ndo13:204107}\citeM{ndo12:034107}. The correctness of the implementation has been checked on many systems by comparing the KEOs with those obtained numerically with the TNUM software \citeM{lau02:8560,tana-tnum:package}. We thus obtain an exact operator. TANA provides the operator in an ascii file that can be directly read by MCTDH. One advantage of the family of polyspherical coordinates is that it always leads to an operator in a sum of products of one-dimensional operators. In the present case,  with those coordinates and their corresponding ranges, we avoid all the possible singularities in the KEO so that we do not need to use 2D DVRs that are numerically less efficient  than products of 1D DVRs.

\subsection*{Sum-of-products of potential and dipole moment surfaces}

In the present case, the potential energy  and dipole surfaces 
were made available to us in the form of a numerical library by 
Joel Bowman and coworkers\citeM{qu18:151,hein18:151}. The potential 
and dipole routines take a single coordinate vector as input and 
return the respective energy value or 3-component dipole vector.

The Heidelberg MCTDH
implementation\citeM{mey90:73,man92:3199,bec00:1,mey03:251,mey09:book,mey12:351} 
relies on an explicit numerical representation of the potential in terms of a
sum of products of one- or low-dimensional functions which are sampled on 
a primitive grid. Hence, given a numerical library routine for the 
potential (and dipoles), a preprocessing step is necessary that creates 
the required numerical representation of the potential from the output of the 
library routines.

In the present case the potential energy surface has been decomposed into a sum-of-products of 2048
low-dimensional terms, more precisely into a Canonical Polyadic Decomposition form. The low-dimensional
basis functions are defined on the coordinates that correspond to those of the bottom layer of the wavefunction tree, 
(cf. Fig.\ S10, extended data).  Such a decomposition
can be used within  the Heidelberg MCTDH package. The decomposition was created 
using a Monte-Carlo variant\cite{sch20:024108} of the alternating least squares
algorithm that is often employed to obtain canonical decompositions. In total eight
symmetries have been incorporated into the PES fit, all of them with respect of 
rotations of the outer water ligands. Other symmetries could not be implemented due to mixing of 
coordinates. For details about the algorithm the reader is referred to Ref.\ \citenum{sch20:024108}. 

The surface fit needs to be performed in the internal dynamical coordinates, 
the library routines usually require Cartesian coordinates to calculate the respective 
potential energy such that here we interlinked the TANA program with fitting program 
to be able to transform between the two sets of coordinates.

%\bibliographystyleM{naturemag}
%\bibliographyM{refs}

\subsection*{Code availability}
The TANA and MCTDH codes including adaptions and input files for the current contribution is available upon request from the authors.

\subsection*{Data availability}
Data used in this contribution is available upon request from the authors.

%\bibliographystyle{naturemag}
%\bibliography{refs}

\subsection*{Acknowledgments}
The authors thank Prof. Mark Johnson and Prof. Knut Asmis for sharing 
experimental data with us and Prof. Joel Bowman for the source code of the
potential energy surface. We furthermore thank the High Performance Computing
Center in Stuttgart (HLRS) under the grant number HDQM\_MCT as well as the bwHPC
project of the state of Baden-W\"urttemberg under grant number bw18K011 for
providing computational resources. The authors thank the CNRS
International Research Network (IRN) "MCTDH" for financial support. 

\subsection*{Author contributions}
M.S., H.-D.M. and O.V. conceived the idea and planed the calculations and the
analysis methodology.
M.S. contributed the SOP fitting of the PES and DMS and performed the dynamical
calculations and analysis.
F.G., D.L. and O.V. designed the coordinates system. F.G. and D.L.  generated
the corresponding analytical KEO.
The text was initially composed by M.S. and O.V., and all authors contributed to
the discussion and interpretation of the results and to the final version of the
manuscript.

\subsection*{Competing interests}
The authors have no competing interests.

\subsection*{Materials \& Correspondence}
Correspondence should be directed to M.S
(markus.schroeder@pci.uni-heidelberg.de) or O.V.
(oriol.vendrell@uni-heidelberg.de).

\subsection*{Supplementary material}
Supplementary material is available for this paper.

\end{document}

\maketitle 
\thispagestyle{empty}

\normalsize{$^\ast$Correspondence: markus.schroeder@pci.uni-heidelberg.de; oriol.vendrell@uni-heidelberg.de}

\vspace{3cm}

This PDF file includes:\\
\hspace*{1cm} Supplementary text\\
\hspace*{1cm} Figs. S1 to S10\\
\hspace*{1cm} Tables S1 to S4\\
\hspace*{1cm} References
\newpage

\setcounter{page}{1}

\section{IR spectrum of the Eigen cation}

\begin{figure}[!ht]
  \centering
  \includegraphics[width=\textwidth]{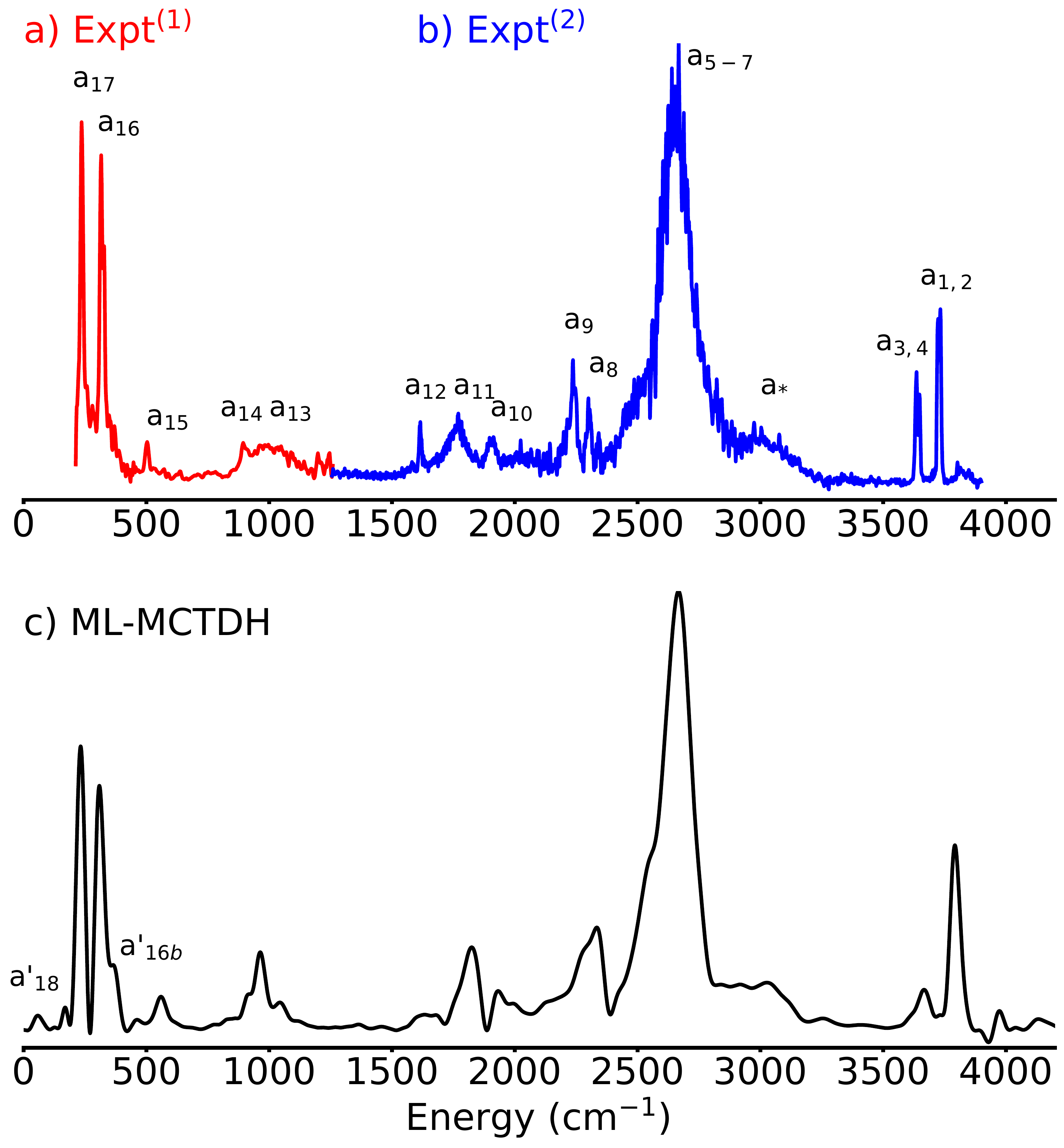}
  \caption{Absorption spectrum of the Eigen Cation H$_9$O$_4^+$. 
  a) Experimental spectrum from Ref.\ \citenum{ess18:798}, 
  b) Experimental spectrum from Ref.\ \citenum{wol16:1131}, 
  c) Calculated spectrum (red-shifted 70 cm$^{-1}$ to match experimental line positions)
  \label{fig:hspectrum}}
\end{figure}

Figure\ \ref{fig:hspectrum} shows the calculated absorption spectrum of the
perprotio Eigen cation
%using a full-dimensional wavefunction and Hamiltonian of the Eigen cation
in comparison with the experimental spectra from Refs.\ \citenum{wol16:1131}
and\ \citenum{ess18:798}.
The overall agreement of calculated and experimental spectra is very good.
The calculated spectrum contains a global shift of 70 cm$^{-1}$ (about
2~cm$^{-1}$ per mode) towards lower energies to better match the main measured
bands.
%experimental line positions, in particular the two dominant peaks a$_{17}$ and a$_{16}$ below 500 cm$^{-1}$ and the large feature a$_{5-7}$ at 2700  cm$^{-1}$.
%
%A global shift  
%we had to include is likely a consequence of a not fully converged wavefunction used in the dynamics. 
The resolution of the calculated spectrum is approximately 30 cm$^{-1}$ and
limited by the 1~picosecond duration of the dipole-dipole correlation function.
Longer correlation times require more tightly converged wavefunctions, which
become prohibitive computationally on current hardware for such a large system. 
As a consequence, the peak positions are considered accurate to a few tens of
cm$^{-1}$ and are rounded to 10 cm$^{-1}$ when listed or discussed.
The calculated peak positions are listed in Table \ref{tab:assign_h} alongside
with experimental results and assignments. 

\begin{table}[!ht]
\centering
 \caption{Peak positions, labels and assignments. Assignments of primed labels
 this work, other assignments from Refs. \citenum{wol16:1131,
 yu17:121102, yu17:10984, ess18:798}. In column 2, we state most important zero
 order assignments with number of quanta and symmetry in the three arms (ABC),
 where
 (+++) represents $\left|100\right> + \left|010\right> +\left|001\right>$,
 (-++) represents $-2\left|100\right> + \left|010\right> +\left|001\right>$  and
 (0+-) represents $\left|010\right> -\left|001\right>$. 
(-++) and (0+-) are degenerate in the S$3$ permutation symmetry group. The coordinates, second column, are explained below.
 All energies in cm$^{-1}$. The MCTDH results include a
 red-shift of 70 cm$^{-1}$ to match peaks a$_{16}$ and a$_{17}$. $^{*}$) from cross-corrlation analysis.}
  \label{tab:assign_h}
 \begin{tabular}{lr lrrr}
 \multicolumn{3}{c}{} & \multicolumn{3}{c}{Peak positions} \\
   \multicolumn{2}{c}{Assignment} & Label & Expt\cite{ess18:798} &  Expt\cite{wol16:1131} & MCTDH  \\
  \hline
 Global O$_4$ pyramidalization         & $\theta_{1,(+++)}$      & a$'_{18}$  &      &      &   60  \\
 Water ligand wagging                  & $\lambda_{1,(-++)}$      & a$_{17}$  & 236  &      &  240  \\
 O-O stretching                        & $R_{1,(-++)}$      & a$_{16}$        & 316  &      &  310  \\
 Water ligand rocking                  & $\beta_{1,(+++)}$      & a$'_{16}$   &      &      &  370  \\
 Water ligand wagging overtone         & $\lambda_{2,(-++)}$      & a$_{15}$  & 503  &      &  560  \\
 Hydronium wagging                     & $y_{1,(-++)}$      & a$_{14}$        & 894  &      &  920  \\
 Hydronium umbrella                    & $y_{1,(+++)}$      & a$_{13}$        & 1014 &      &  970  \\
 OH-bend water ligands                       & $b_{1,(-++)}$      & a$_{12}$  & 1615 & 1615 & 1620$^{*}$  \\
 Hydronium wagging overtone            & $y_{2,(-++)}$      & a$_{11}$        & 1760 & 1770 & 1830  \\
 Hydronium umbrella + wagging          &                    & a$_{10}$        &      & 1909 & 1940  \\
 OH-bend + frustrated hydronium rotation &                  & a$_{9}$         &      & 2237 & 2290  \\
 OH-bend + hydronium umbrella          &                    & a$_{8}$         &      & 2300 & 2340  \\  
 Proton transfer                       & $z_{1,(-++)}$      & a$_{5-7}$       &      & 2653 & 2670  \\
 Comb. O-O stretch and proton transfer & $R_1z_{1,(-++)}$   &                 &      &      &       \\
 Symmetric OH- stretch water ligands   & $v^{(s)}_{1,(-++)}$& a$_{3,4}$       &      & 3636 & 3670  \\
 Asymmetric OH-stretch water ligands   &$v^{(a)}_{1,(-++)}$ & a$_{1,2}$       &      & 3724 & 3800  
 \end{tabular}
 
\end{table}

Figure\ \ref{fig:hspectrum} and Table \ref{tab:assign_h} indicate the assignments of the
main peaks, where we adhere to the labeling convention in Ref.\ \citenum{wol16:1131}.
The peaks labelled on top of the experimental spectra have been discussed
in Refs. \citenum{wol16:1131, yu17:121102, yu17:10984, ess18:798}, whereas peaks labelled
on top of the calculated spectrum correspond to new assignments.
The quantum mechanical localized excitations indicated in the assignments represent the
largest contribution to the corresponding eigenstate. The assignment is established
through cross-correlating locally excited test states with the propagated, dipole-operated
ground state. Based on the cross-correlation analysis, the band a$_{11}$ 
 can be unambigously assigned to an overtone of
the hydronium wagging. Previous works had suggested a similar asignment without detailed
reference to specific overtone and combination band~\cite{yu17:121102}.

%bands a$_{10}$ and a$_{11}$
%we can be unambigously assigned to specific combinations of
%the hydronoum umbrella and wagging motion, and to an overtone of
%the hydronium wagging, respectively.
%Previous works had suggested similar asignments without detailed
%reference to specific overtones and combination bands~\cite{yu17:121102}.

Our analysis provides further assignments including
peaks a$_1$ -- a$_{5-7}$, and a$_{13}$ -- a$_{17}$ 
(see cross-correlation analysis below). 
The peak a$_{12}$ at 1615  cm$^{-1}$ in the experiment which 
belongs to O-H bending is blurred in the calculated spectrum but is found in the cross-correlation analysis at 1620 cm$^{-1}$ (see cross-correlation analysis below).
We have confirmed, based on separate calculations using a different multilayer
tree, that this blurring is due to a lack of convergence of the propagated
wavefunction. 
The subtree of the hydronium core is especially challenging to converge and highlights the
very strong coupling of the central unit with its solvation shell.
This very strong coupling results in possibly the least well understood
feature of the gas-phase spectrum of the Eigen cation, namely the
very broad absorption band centered at 2600~cm$^{-1}$ in Fig.~\ref{fig:hspectrum}.
%
%For peaks a$_8$ to a$_{10}$ appropriate zero-order test states could not be created with satisfactory 
%precision as the set of used internal coordinates do not facilitate generation of these states.

%\subsection{Anharmonic couplings between the excess proton and its first solvation shell}

Similarly to the much better understood Fermi-resonance double peak in the
Zundel cation~\cite{ven07:6918, ven09:352}, this very broad band has its origin
in the coupling of the proton transfer modes in the Eigen cation to other
vibrational modes of the system~\cite{wol16:1131}.
The agreement of the width and shape of this band between the calculated and
experimental spectra is excellent. The spectral resolution of the calculation is
about 10 times narrower than the calculated band's width, indicating that this
is a genuine characteristic of the band and not caused by experimental
artifacts, e.g., due to the tagging agent. The pre-band features a$_{11}$ --
a$_{8}$ and post-band feature a$_*$ are all reproduced as well.

\begin{figure}[!ht]
  \centering
  \includegraphics[width=0.8\textwidth]{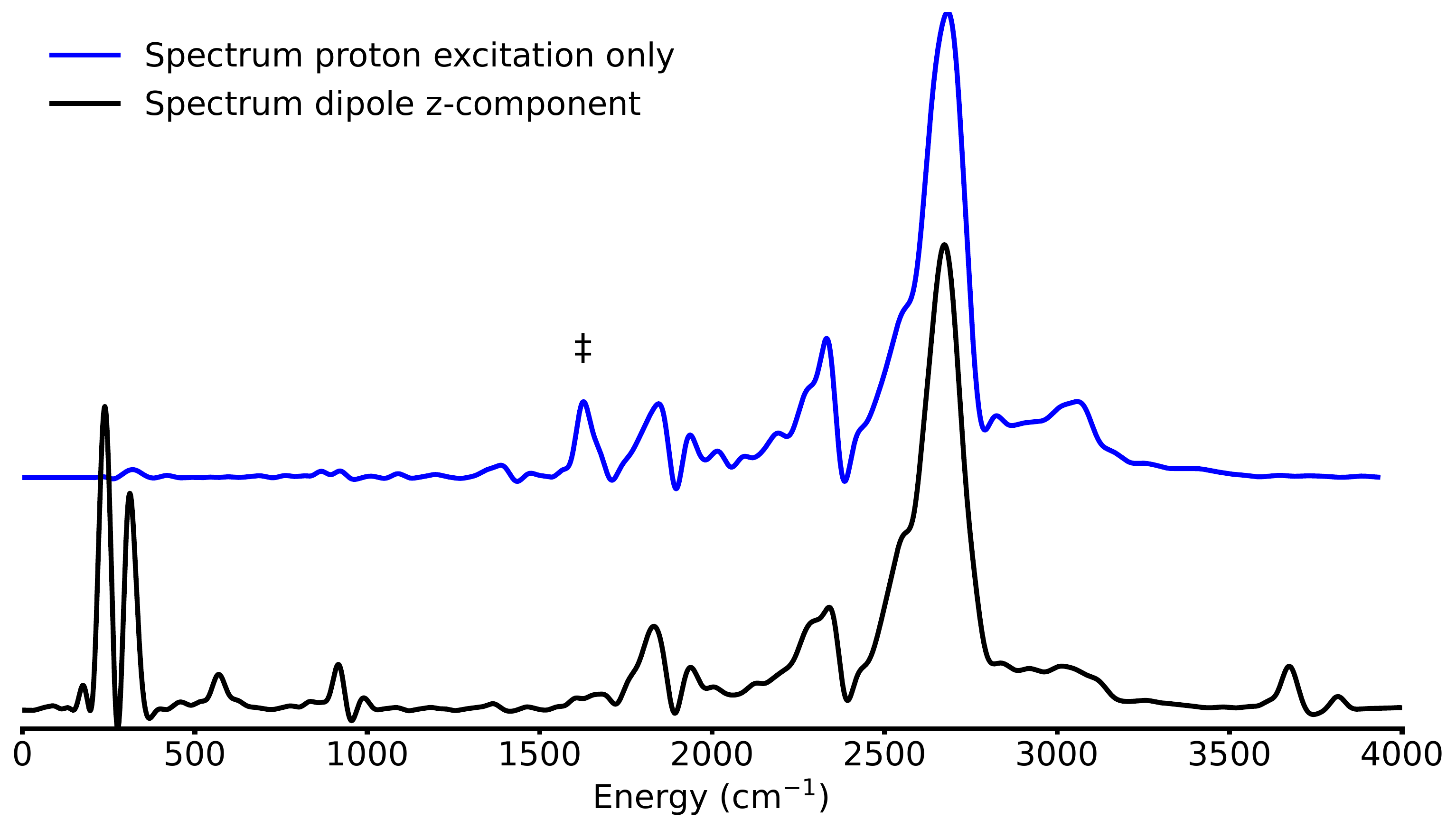}
  \caption{Spectrum obtained with a model dipole exclusively operating on proton
  transfer coordinates (blue curve) in comparison with the spectrum of the full
  dipole z-component (black curve).  Excitation of the proton transfer excites
  multiple other states and leads to a characteristically broad proton transfer
  signal. The peak marked with $\ddag$ at 1620  cm$^{-1}$, blurred in the 
  dipole-spectrum, corresponds to O-H bending of the ligand water molecules.
  \label{fig:spec_zmpp}}
\end{figure}

\begin{figure}[!ht]
  \centering
  \includegraphics[width=0.5\textwidth]{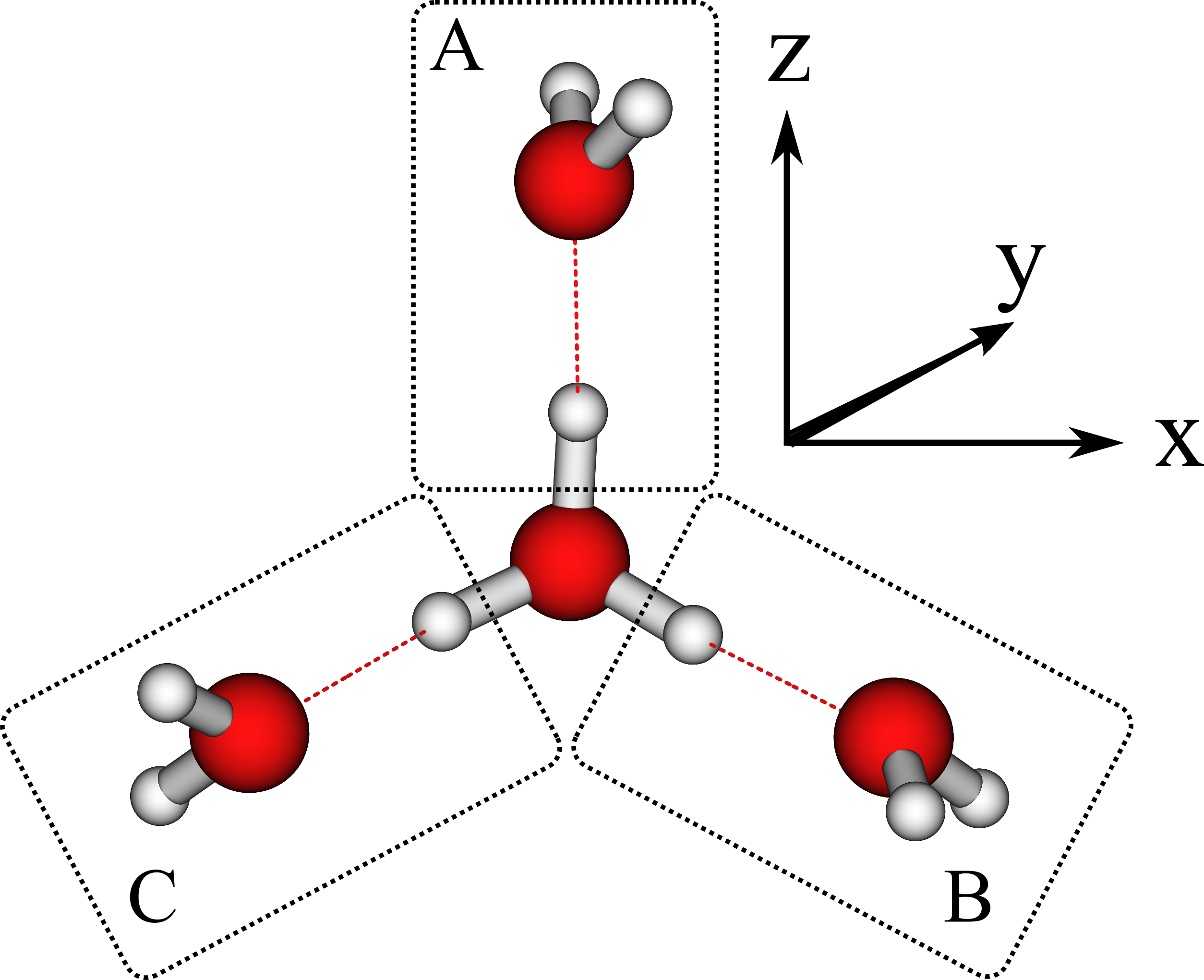}
  \caption{Labeling of the three "arms" of the eigen cation as "A", "B" and "C" as well body fixed axis and
  directions of dipole moment components. The z-component points along the O-O vector of arm "A".}
  \label{fig:eigenabc}
\end{figure}

The localized excitation of the proton transfer coordinates of the cation with a
model dipole operator $\mu_{\rm test}=-2z_{\rm A} +z_{\rm B} + z_{\rm C}$
results in a spectrum with all fundamental spectral features of the shared
proton modes. This operator has the same symmetry as the $z$-component of the
dipole operator, but it exclusively operates on the three proton transfer
coordinates $z_{\rm X=A,B,C}$. The coordinate $z_{\rm X}$ describes the proton motion
parallel to the corresponding O-O vector.
 The coordinate system is explained in Figures \ref{fig:eigenabc} and \ref{fig:coord}.
As seen in Fig.\ \ref{fig:spec_zmpp} (blue curve), the bands related to the
proton motion span about 1500~cm$^{-1}$, and the main proton transfer feature is
nearly identical to the one in the full spectrum (black curve).
The character of the various peaks flanking the broad proton transfer band is
listed in Table~\ref{tab:assign_h} and comprises excitations of the hydronium
core, oxygen-oxygen stretches and water bending motions in the first solvation
shell.

Using reduced dimensional models we explain the pronounced broadening of the
proton transfer peak by coupling with many other states.
Combinations of low excitations in the hydronium core and  excited states in O-O
- distance and ligand wagging modes play an important role in this broadening
process.  The involvement of wagging states and O-O distance can be seen as an
analogy to the Zundel cation, where a Fermi resonance between one excitation of
the proton transfer and a combination mode of the O-O distance mode and a
wagging overtone exists.  As opposed to the Zundel cation, where only one
combination excitation broadens/splits the proton transfer line, there are many
in the Eigen cation.

\subsection{Spectral components}
\begin{figure}[htbp]
  \centering
  \includegraphics[width=0.8\textwidth]{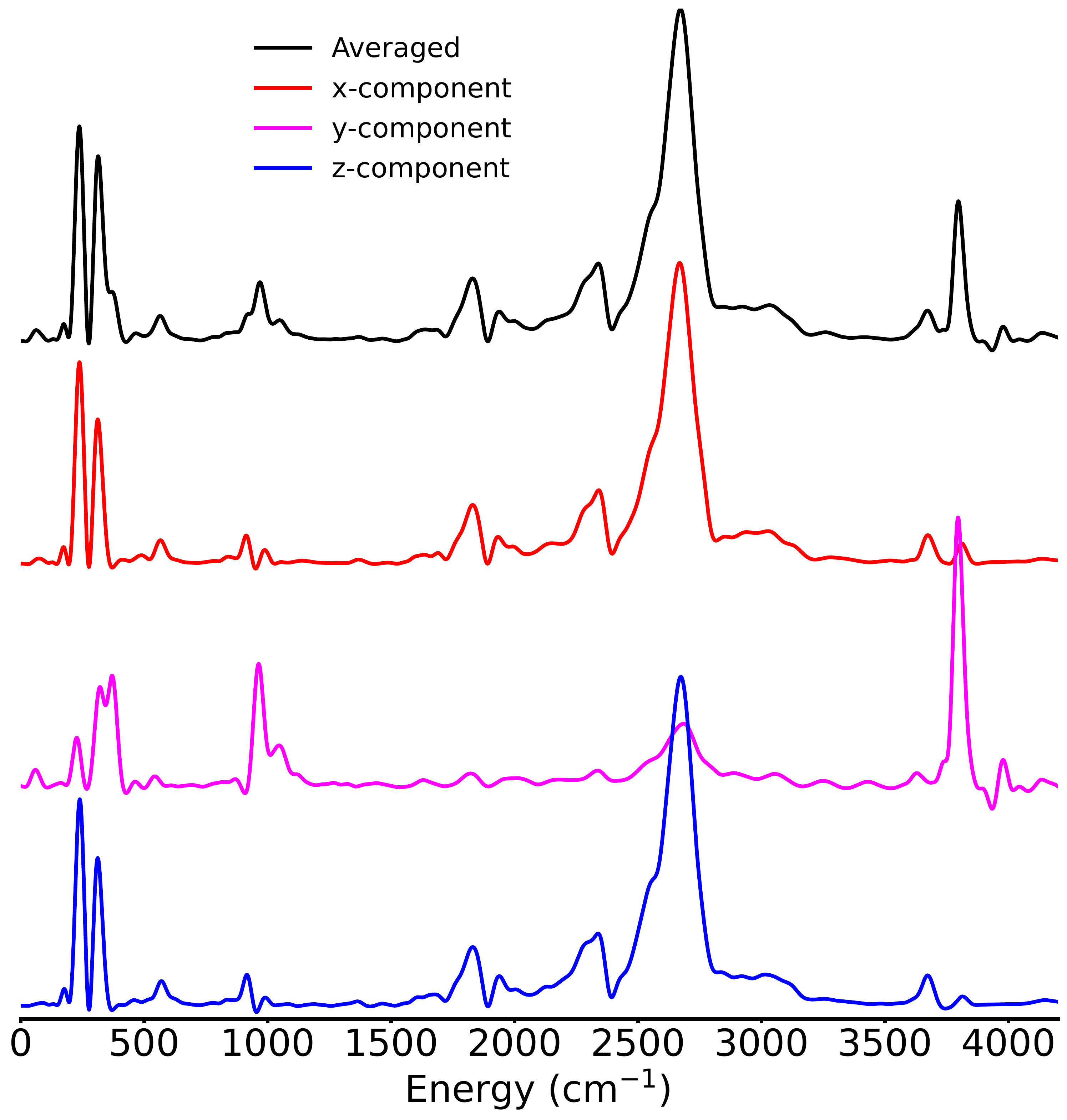}
  \caption{Calculated averaged absorption spectrum (upper panel) and $x-$, $y-$ and $z-$ components (lower panel) of the Eigen cation, H$_9$O$_4^+$. Relative intensities of the components 
  according to their contribution to the averaged spectrum.
  \label{fig:spectrum_components}}
\end{figure}

Fig. \ref{fig:spectrum_components} shows the averaged calculated spectrum (black curve) 
and the spectra obtained with the dipole constituents $\mu_x$ (red curve), $\mu_y$ (green curve) and $\mu_z$ (blue curve) for H$_9$O$_4$. 
The single spectra are to scale, i.e., the single dipole component spectra are displayed according to their 
contribution to the averaged spectrum.

\section{Assignments}
We restrict our discussion below to the cases 
$i=z$ (in-plane, parallel to one hydrogen bond of the molecule) and $i=y$ (out-of-plane) as the states for $i=x$ 
(remaining in-plane)  resemble the features of the other in-plane dipole component $i=z$.

\begin{figure}[!ht]
  \centering
  \includegraphics[height=.82\textheight]{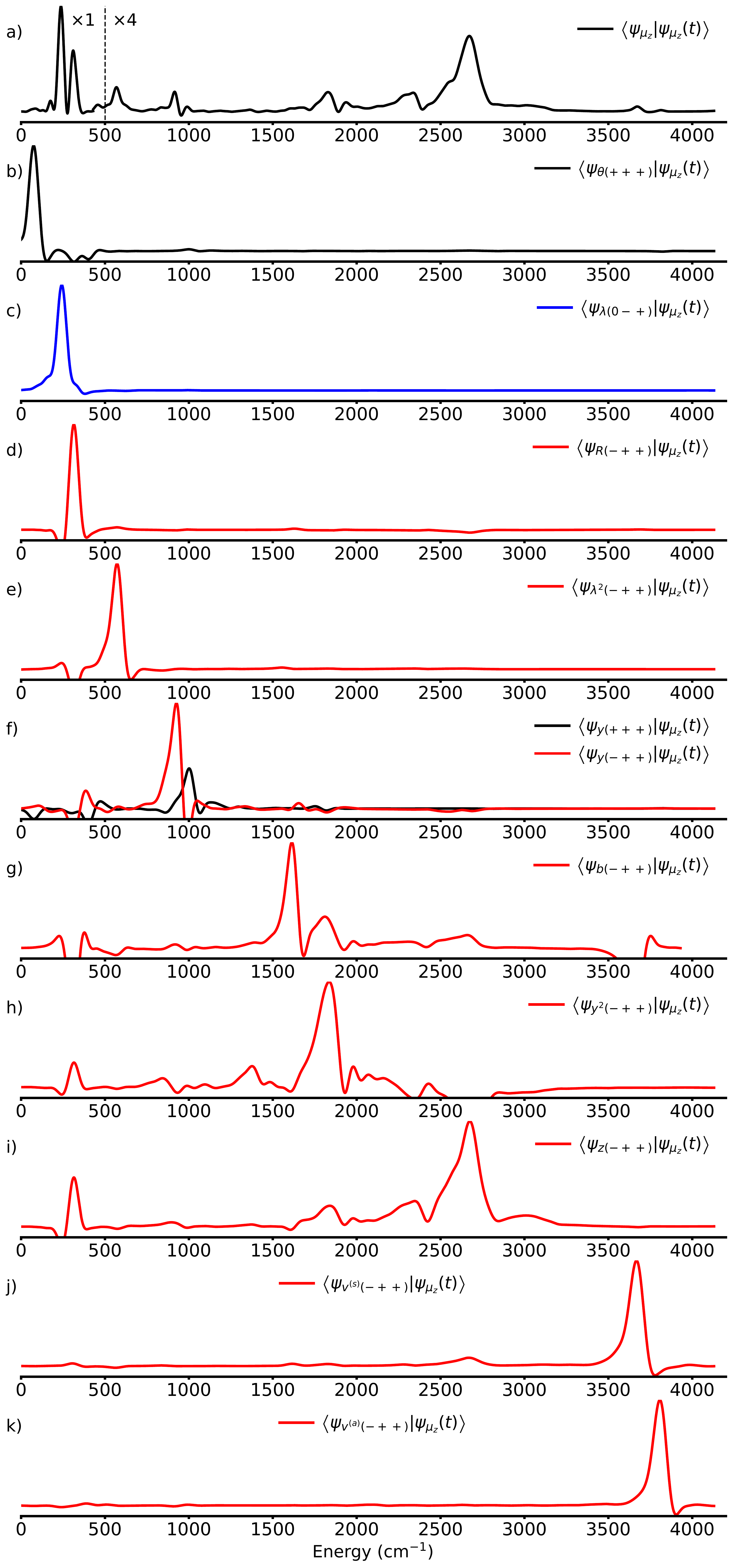}
  \caption{Fourier transforms of cross-correlation functions between the 
  time-dependent $\mu_z$-operated ground state of H$_9$O$_4^+$ and test states. The labels show the definition of the 
  correlation function. The curves are scaled to a maximum 
  value of unity. Where more than one curve is shown in one panel their relative weights are to scale. 
  Panel a) shows the Fourier transform of the auto-correlation function 
   scaled by a factor of four above 600 cm$^{-1}$.
  \label{fig:cross_spec_dipz}}
\end{figure}

In Fig.\ \ref{fig:cross_spec_dipz} the Fourier transforms of the cross-correlations 
of test states with the time-propagated dipole-operated ground state for the
z-component of the dipole are given in panels b-k) while panel a) shows the Fourier 
transformed auto-correlation function (scaled by factor 4 above 500 cm$^{-1}$).  
The peak heights of the auto-spectrum hence correspond to the 
population of the respective eigenstates in the dipole-operated ground state.
Note that the labels in Fig. \ref{fig:cross_spec_dipz} denote the correlation function while
the graphs show their Fourier transform without the $\omega$ prefactor, Eq.\ (5) of Methods.
The curves are color-coded for different liner combinations of coordinate operators: black (+++),
red (-++) and blue (0+-). 

In panels b-e) the cross-correlations for the b) global O$_4$ umbrella motion, c) 
the asymmetric wagging of the outer water molecules, the d) asymmetric O-O stretch and
e) the asymmetric double excited wagging signals
are shown, respectively. Other liner combinations than the one displayed 
have not been found to have significant intensity. 

In panel f) the symmetric umbrella (black) and
asymmetric out-of plane wagging (red) of the central hydronium H-atoms are probed.
The two curves are to scale, i.e., the relative peak height corresponds to the 
relative intensity of both signals.

The ligand water O-H-bending signal, which is blurred in the calculated spectrum of H$_9$O$_4$, can be clearly identified in the cross correlation analysis, panel g). The bending is modeled as a linear combination of the internal Jacobi vectors of the waters. We believe that the blurring is due to the relatively small basis set used here.

Panel h) of Fig.\ \ref{fig:cross_spec_dipz} shows the cross-correlation signal of the 
doubly excited asymmetric umbrella motion of the central hydronium H-atoms. The main contribution
of this signal lies slightly below 2000 cm$^{-1}$ matching the respective peak in the absorption spectrum.
 
In panel i) the asymmetric O-H of the central hydronium are probed. One clearly sees that
the broad feature in at 2700 cm$^{-1}$  corresponds to the proton transfer which is correlated with many 
other states in particular between 1600 
and 3500 cm$^{-1}$. Also the O-O distance state, which appears near 450 cm$^{-1}$ gains significant intensity.

Finally in panel j) and k) the symmetric and asymmetric  O-H-stretching of the outer water ligands are probed. For the symmetric  O-H-stretching a linear combination of the internal Jacobi vectors of 
the water ligands has been used.

Furthermore, for 
the remaining states of the hydronium core  
above 2000 cm$^{-1}$, it proved difficult to create appropriate test states such that these signals could not be unambiguously assigned.

\begin{figure}[!ht]
  \centering
  \includegraphics[height=.82\textheight]{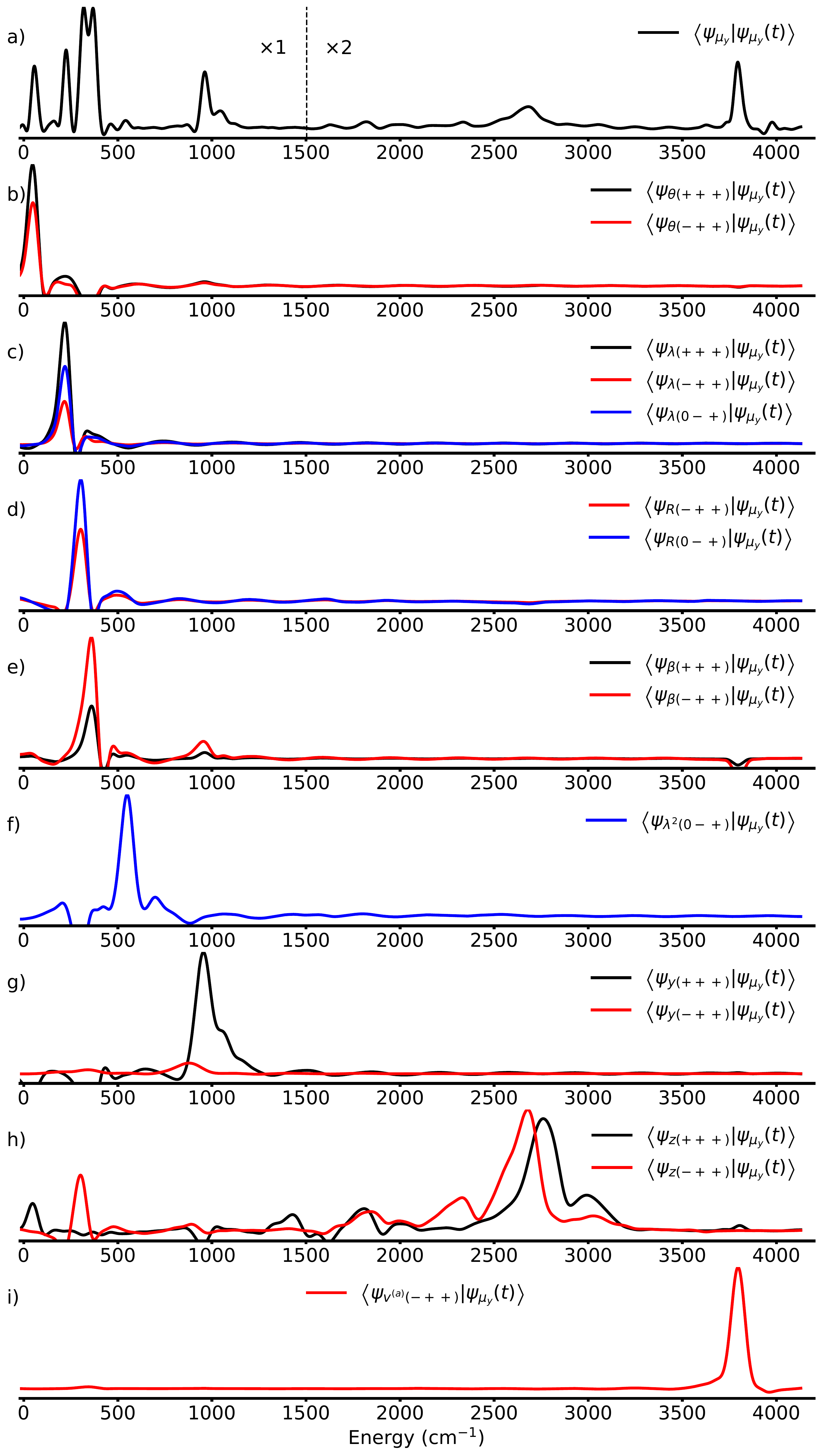}
  \caption{Fourier transforms of cross-correlation functions between the 
  time-dependent $\mu_y$-operated ground state of H$_9$O$_4^+$ and test states (details see text)
  to probe existence of a corresponding state.  The curves are scaled at maximum 
  to unity. Where more than one curve is shown in one panel their relative weights are to scale. 
  The upper panel shows the Fourier transform of the auto-correlation function scaled by
  a factor of two above 1500 cm$^{-1}$.
  \label{fig:cross_spec_dipy}}
\end{figure}

In Fig. \ref{fig:cross_spec_dipy}, the Fourier transforms of the cross-correlations of test states with the time-propagated dipole-operated ground state for the y-component of the dipole are given. 
Panel a) again shows the auto-correlation signal of the $\mu_z$-operated ground state. The signal is enlarged by a factor of 
two above 1500 cm$^{-1}$. 
The structure in the lower frequency range up to 1000 cm$^{-1}$ is much more pronounced than for the 
$z$-component. Panels b-f) show the respective 
cross-correlation signals of the b) O$_4$ global angles, c)  wagging of the outer water ligands, d)  O-O stretch, e)  H$_2$O ligand rocking, and f) asymmetric wagging overtone. 

In panel g) the asymmetric wagging (black curve) and symmetric umbrella (red curve) signals of the central hydronium are shown. The feature at 1000 cm$^{-1}$ is mainly caused by the umbrella motion 
and a second peak on the high energy shoulder of the umbrella motion is visible. 

Panel h) shows the cross-correlation signals with
the symmetric and asymmetric O-H-stretching modes of the
central H$_3$O$^+$. Both peaks contribute approximately equally to the signal in the y-component of the dipole. Note, that the peaks 
are energetically quite separated but equally broad. Excitation of the proton transfer hence seems to co-excite quite different states 
in the energetic vicinity of the original excitation. Note that in case of the symmetric O-H-stretching also the global O$_4$ scaffold is affected. 
The asymmetric O-H-stretching of the outer water ligands is finally probed in panel i).

\section{Transition from Zundel@Eigen to Zundel}

\begin{figure}[!ht]
  \centering
  \includegraphics[width=0.8\textwidth]{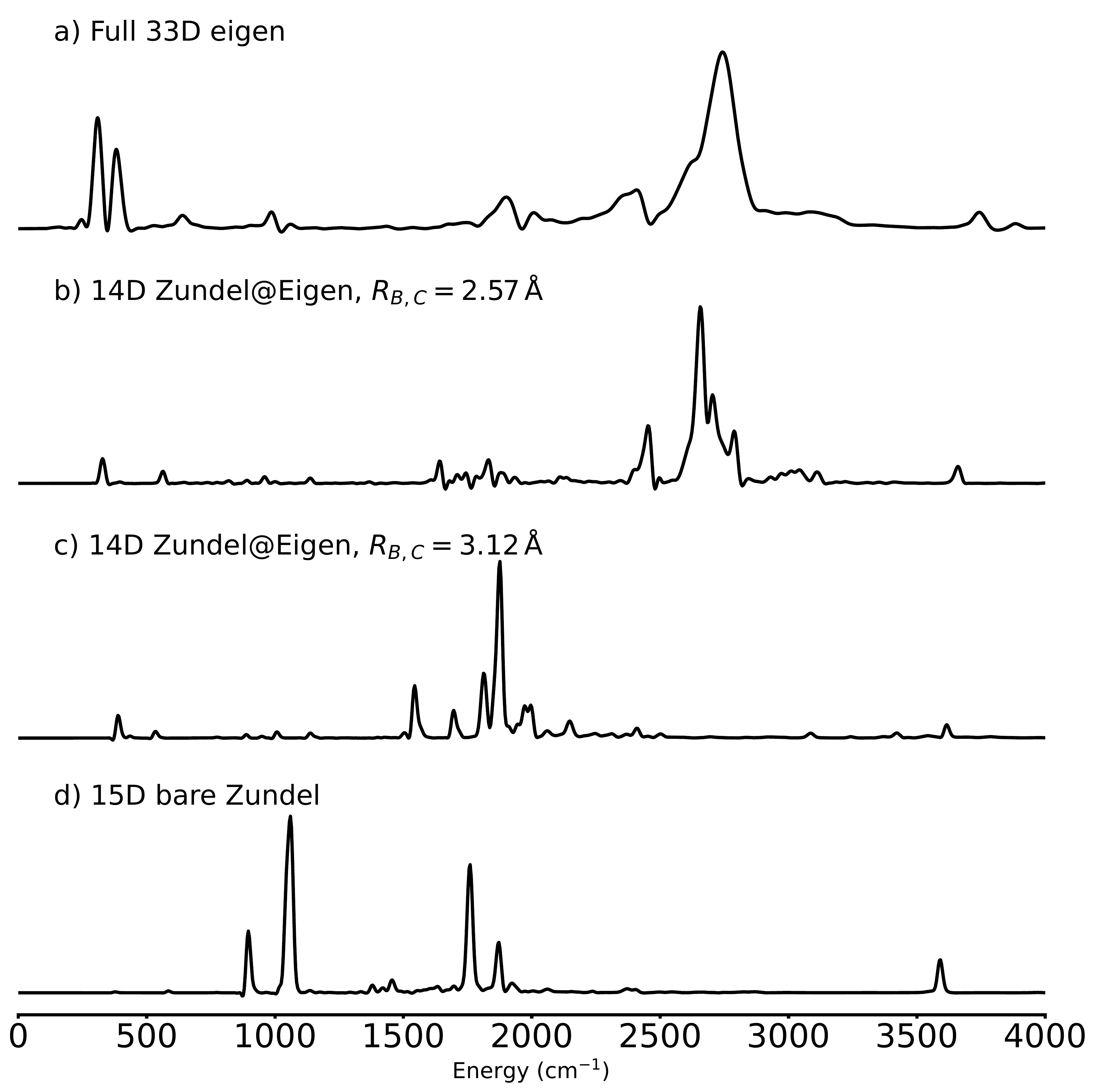}
  \caption{Spectra obtained with the z-component of the dipole moment surfaces for various systens:
  a) Full 33D calculation of the eigen cation 
  b) 14D calculation of the zundel subunit as in Fig. 4a) of the main text with the two frozen O-O distances set to equilibrium positions of 2.57 \AA. 
  c) Same as in b) but with the two frozen O-O distances set to 3.12 \AA. 
  d) Full 15D Zundel using the PES from Ref. \citenum{hua05:044308}.
  \label{fig:distant_OO}}
\end{figure}

To justify the argumentation of the Zundel subunit being the fundamental dynamical building block to understand the spectrum of the Eigen cation
we calculated spectra of various configurations as depicted in Fig.\ \ref{fig:distant_OO}.  It illustrates the influence of the presence of the static Eigen environment on the position of proton transfer peak. 

 Fig.\ \ref{fig:distant_OO} a) 
shows the spectrum obtained with the dipole moment $z$-component using 
a full-dimensianal quantum mechanical model with the proton transfer peak visible at 2700 cm$^{-1}$. In Fig. Fig.\ \ref{fig:distant_OO} b) only one Zundel subunit of the total Eigen
cation is modeled dynamically while all other coordinates are frozen to
their equilibrium position. One observes that the spectrum changes slightly, but the broad proton transfer remains approximately at 2700
cm$^{-1}$. 

In Fig.\ \ref{fig:distant_OO} c) the two frozen water molecules 
are moved away from the dynamical zundel subunit by increasing the
respective O-O distance by 0.55 \AA\phantom{a}to 3.12 \AA. This has dramatic consequences for the spectrum as the proton transfer peak is now red-shifted by about 900 cm$^{-1}$. 

The PES of Refs. \citenum{qu18:151,hein18:151} does not facilitate 
larger O-O distances hence Fig.\ \ref{fig:distant_OO} d) shows the spectrum of the bare Zundel
cation for comparison. This spectrum has been obtained using the 
PES and DMS of Huang {\it et al.}\cite{hua05:044308}.  The proton transfer peak here is located at slightly above 1000 cm$^{-1}$, hence red-shifted by about 1600 cm$^{-1}$ compared to Eigen.

\newpage

\subsection*{Kinetic energy operator}

\begin{figure}[!ht]
  \centering
  \includegraphics[width=0.6\textwidth]{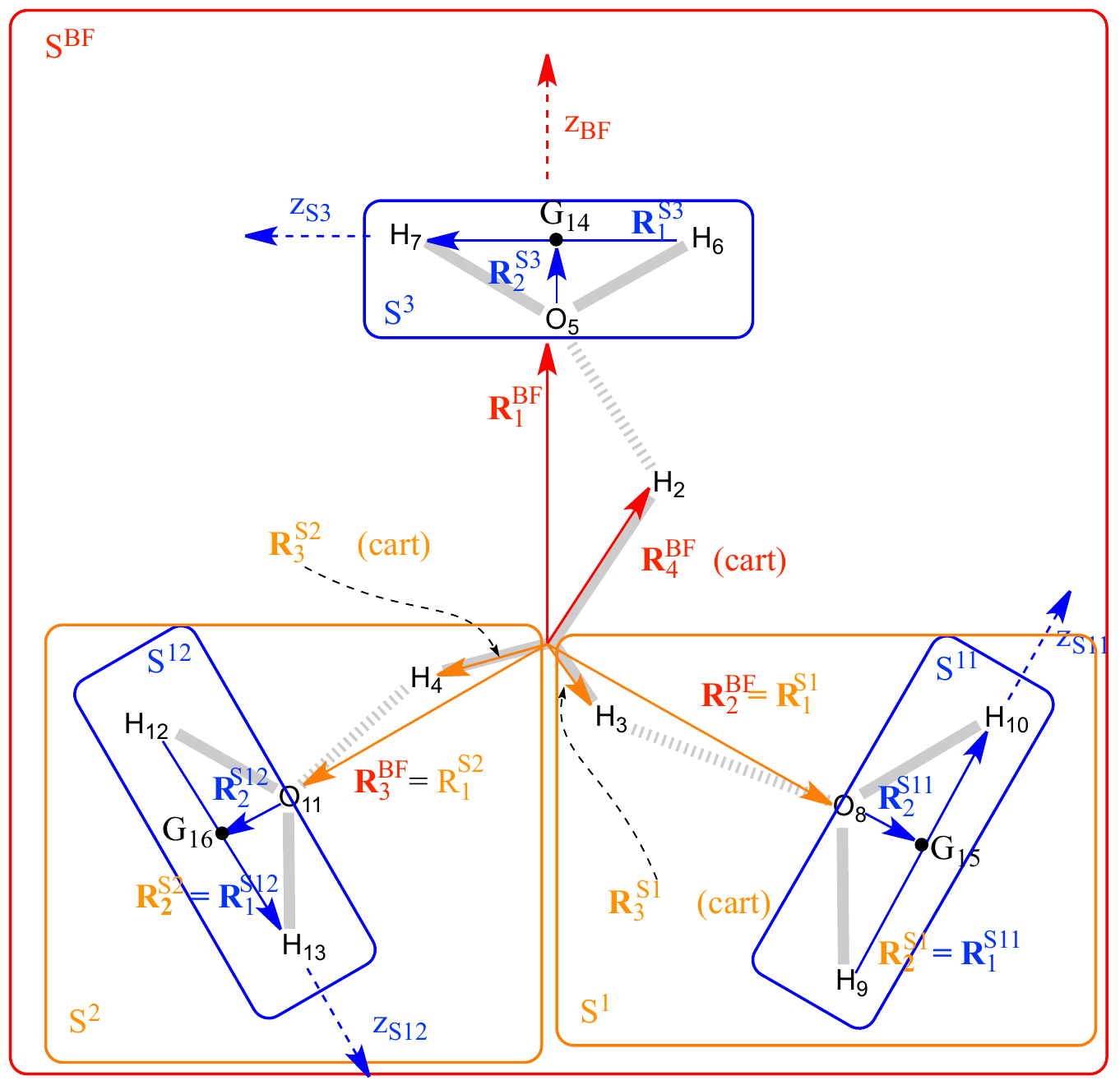}
  \caption{Structure of the vectors for the description in polyspherical coordinates. Two Jacobi vectors (one from one H atom to the other and the second from
  the O atom to the middle of H$_2$) are used to describe the molecules of water (in blue).
  \label{fig:vectors-polyspher}}
\end{figure}

\subsection*{Coordinates and primitive basis \label{sec:numsetup}}

\begin{figure}[!ht]
  \centering
  \includegraphics[width=\textwidth]{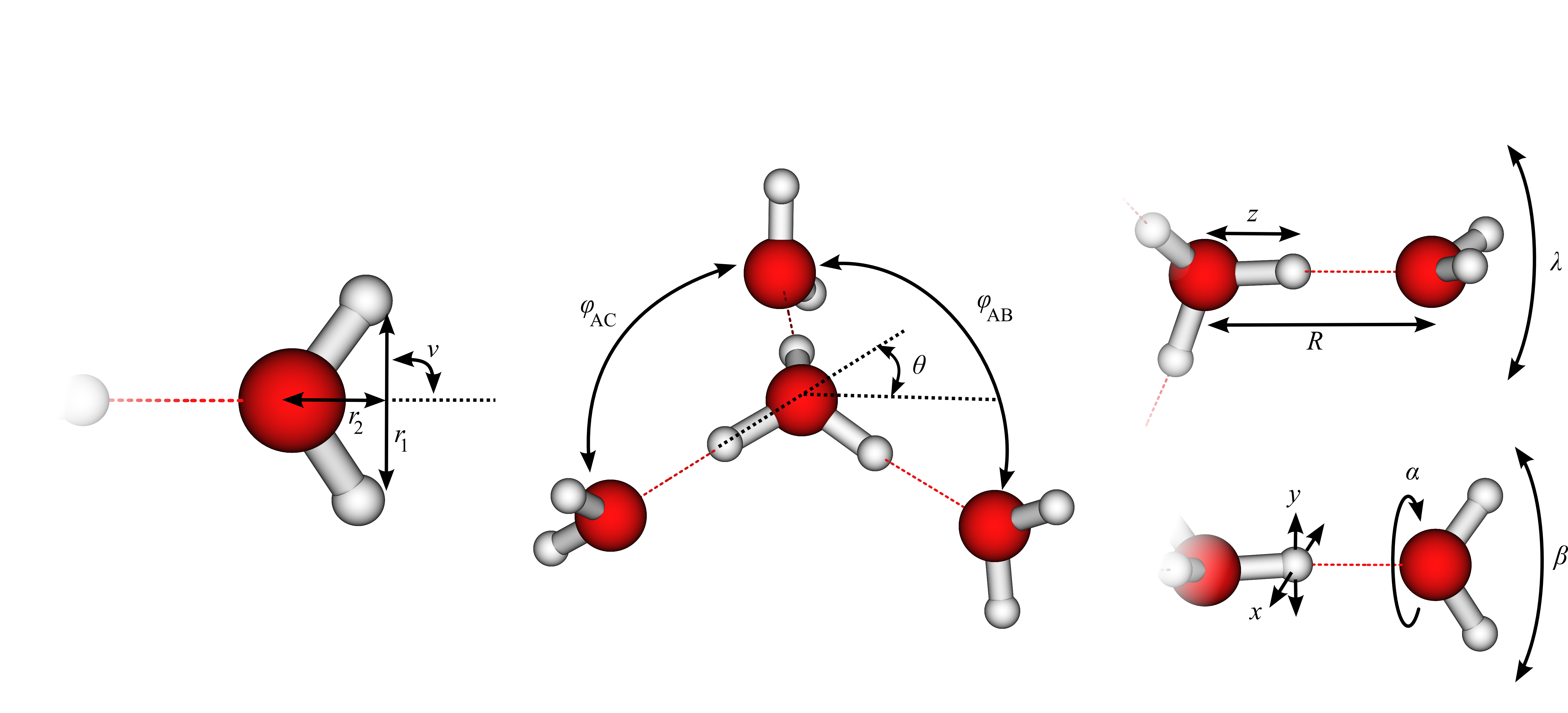}
  \caption{Coordinates used for dynamical calculations. Coordinates defined in one "arm" of the cluster are additionally 
  labeled with subscript A, B or C, depending in which arm they are defined. The three arms are labeled clockwise A, B and C.  
  \label{fig:coord}}
\end{figure}

Table \ref{tab:primitive} outlines the 
physical coordinates, coordinate ranges and  number of primitive grid points used for each coordinate. 
A graphical illustration of the coordinates is also given in Fig. \ref{fig:coord}.

\begin{table*}[!ht]
\caption{Primitive coordinates and grids.} \label{tab:primitive}
\resizebox{\textwidth}{!}{%
\begin{tabular}{rrrrcl}
 Coordinate label & physical meaning&  DVR type & No. Points & range &\\
 \hline
 $\theta$                     & O$_4$ pyramidalization & sin &  11  & 2.142 --  3.142 & rad \\
 $\cos\left(\varphi_{\rm AB}\right)$,  $\cos\left(\varphi_{\rm AC}\right)$  & O$_4$ bending    & sin &  11  & -0.71 -- -0.1  &    \\
 $R_{\rm A}$,  $R_{\rm B}$, $R_{\rm C}$  &  O-O stretching     & ho  &  13  & 4.3 --  5.9  & au   \\
 $\lambda_{\rm A}$,  $\lambda_{\rm B}$, $\lambda_{\rm C}$ & ligand wagging & sin & 13 &  2.169 -- 4.114  &rad\\  
 $\cos\left(\beta_{\rm A}\right)$, $\cos\left(\beta_{\rm B}\right)$, $\cos\left(\beta_{\rm C}\right)$ & ligand rocking  & sin & 9 &  -0.5 -- 0.5  & \\ 
 $\alpha_{\rm A}$,  $\alpha_{\rm B}$, $\alpha_{\rm C}$ & water rotation around O-O axis  & sin & 9 &  -0.5 -- 0.5  &rad\\ 
 $r_{\rm 1,A}$, $r_{\rm 1,B}$, $r_{\rm 1,C}$ &  water H-H distance      & ho  & 9 & 2.3 --  3.6  & au  \\
 $r_{\rm 2,A}$, $r_{\rm 2,B}$, $r_{\rm 2,C}$ &  water (H-H)-O distance  & ho  & 7 &  0.7 --  1.6  & au  \\
 $\cos\left(\nu_{\rm A}\right)$,  $\cos\left(\nu_{\rm B}\right)$, $\cos\left(\nu_{\rm C}\right)$ & water Jacobi angle & ho & 7  &  -0.3 --  0.3 &\\
 $x_{\rm A}$,  $x_{\rm B}$, $x_{\rm C}$  & hydronium H in-plane     & ho   & 9   &  -0.84 -- 0.84  & au  \\
 $y_{\rm A}$,  $y_{\rm B}$, $y_{\rm C}$  & hydronium H out-of-plane & ho   & 9   &  -0.84 -- 0.84  & au  \\
 $z_{\rm A}$,  $z_{\rm B}$, $z_{\rm C}$  & hydronium O-H-stretching  & ho  & 9   &  1.35 --  2.65  & au  
\end{tabular}
}
\end{table*}


\subsection*{Potential energy and DMS fits}

The potential energy and dipole moment surface fits have been created with in total $7\cdot10^6$ sampling points
which were obtained with a Metropolis algorithm at different temperatures such that the sampling points 
are distributed according to the weight
\begin{equation}
 W(\vec{q}) = \sum_i a_i \exp\left(-\frac{V(\vec{q})}{k_{\rm B}T_i}\right),
\end{equation}
where $T_i$ are the different temperatures and $k_{\rm B}$ is the Boltzmann constant. The coefficient $a_i$ is proportional 
to the number of sampling points used for the given temperature. 
The number of points for each temperature and the root-mean-square error (RMSE) of the fit on a validation set 
obtained with the same temperature but of a fixed size of 10$^7$ points are listed in Table \ref{tab:rmsV}. 
\begin{table*}[!ht]
  \centering
\begin{tabular}{r|c|c}
 \multirow{2}{2cm}{\centering $k_{\rm B}T$ \\in cm$^{-1}$} & No. Points        &  RMSE in cm$^{-1}$ on validation set\\
                          & (creating set)    &  (10$^7$ points each $k_{\rm B}T$) \\
 \hline
 500                      & $2.0\cdot10^6$    &  ~~43\\
 1000                     & $2.0\cdot10^6$    &  ~~77\\
 2000                     & $2.0\cdot10^6$    &  137\\
 3000                     & $7.5\cdot10^5$    &  191\\
 4000                     & $2.5\cdot10^5$    &  244\\
\end{tabular}
\caption{Number of sampling points used for generating the PES fit at different temperatures as well as root-mean-square error of the final potential
on validation sets.} \label{tab:rmsV}
\end{table*}

\begin{table*}[!ht]
  \centering
\begin{tabular}{r|c|c|c}
 $k_{\rm B}T$ in cm$^{-1}$ &  \multicolumn{3}{c}{RMSE in 10$^{-2}$ Debye on validation set}\\
 \hline                   & x-component & y-component & z-component \\
 500                      & 0.5         &  0.6        & 0.7\\
 1000                     & 1.0         &  1.0        & 1.2 \\
 2000                     & 1.9         &  1.7        & 2.3\\
 3000                     & 2.7         &  2.0        & 3.2\\
 4000                     & 3.5         &  2.4        & 4.0\\
\end{tabular}
\caption{Root-mean-square errors of the DMS fits on validation sets with 10$^7$ points}. \label{tab:rmsDMS}.
\end{table*}
A similar procedure has been applied to decompose the $x$, $y$ and $z$ components the
molecular dipole moment surfaces into a sum-of-products form with 1024 terms. Each DMS was 
fitted and validated with the same sets of points as used for the potential.
The RMSE Errors of the fits are listed in Table \ref{tab:rmsDMS}.

\subsection*{Wavefunction representation}

Within MCTDH the wavefunction is represented in a hierarchical Tucker format where a layered set of basis functions is 
used to combine physical into logical coordinates. The ML-representation of wavefunction used in the present case is shown in 
Fig. \ref{fig:mltree}.

\begin{figure}[!ht]
  \centering
  \includegraphics[width=\textwidth]{fig/mltree}
  \caption{Multi-Layer tree used in MCTDH calculations for H$_9$O$_4^+$. Circles denote vertices
  of logical coordinates and rectangles of physical coordinates. Edge labels represent numbers 
  of basis functions. Size of wavefunction tensor: 1.2$\cdot 10^6$ complex numbers. 
  \label{fig:mltree}}
\end{figure}

%\bibliography{refs}
%\bibliographystyle{naturemag}
\input{supporting_eigen.bbl}